\documentclass[twocolumn]{aastex631}

\usepackage{amsmath}
\usepackage{graphicx}
\usepackage{booktabs}
\usepackage{multirow}
\usepackage{xcolor}
\newcommand{\rev}[1]{\ifmmode{\color{blue}#1}\else{\color{blue}\bfseries #1}\fi}
\usepackage{orcidlink}

\begin{document}

\title{{Ranking Prompt--Plateau Correlations in Gamma-Ray Bursts with Symbolic Regression}}
\shorttitle{{Ranking GRB Prompt--Plateau Relations}}

\author{Fatemeh~Rastegarnia}
\affiliation{ICRANet, Piazza della Repubblica 10, 65122 Pescara, Italy}

\author{Rahim Moradi}
\email{rahim.moradi@icranet.org}
\affiliation{ICRANet, Piazza della Repubblica 10, 65122 Pescara, Italy}
\affil{State Key Laboratory of Particle Astrophysics, Institute of High Energy Physics, Chinese Academy of Sciences, Beijing 100049, China}

\author{Brwa Shukur Ismael}
\affil{State Key Laboratory of Particle Astrophysics, Institute of High Energy Physics, Chinese Academy of Sciences, Beijing 100049, China}
\affiliation{University of Chinese Academy of Sciences, Chinese Academy of Sciences, Beijing 100049, People’s Republic of China \\
}
\author{Mohsen Asaadi}
\affil{Yau Mathematical Sciences Center, Tsinghua University, Beijing, 100084, China}

\author{Yu Wang}
\email{yu.wang@icranet.org}
\affiliation{ICRANet, Piazza della Repubblica 10, 65122 Pescara, Italy}
\affiliation{ICRA, Dipartimento di Fisica, Sapienza Universita di Roma, Piazzale Aldo Moro 5, I-00185 Roma, Italy}
\affiliation{INAF -- Osservatorio Astronomico d'Abruzzo, Via M. Maggini snc, I-64100 Teramo, Italy}

\begin{abstract}

{Gamma-ray burst (GRB) correlations probe central engines, relativistic outflows, and radiation processes. The Dainotti relation is an anticorrelation between the X-ray plateau luminosity $L_a$ and rest-frame plateau duration $T_a^{\rm rest}$. We use symbolic regression to compare it with other low-dimensional correlations in 251 GRBs with Swift-XRT plateau properties and Swift-BAT prompt-emission parameters. We perform a Dainotti-guided search, in which $\log T_a^{\rm rest}$ is always present, and an unrestricted search over subsets of at most three variables. The guided search recovers the Dainotti dependence; a D'Agostini fit gives $\log L_a=(51.92\pm0.30)-(1.31\pm0.09)\log T_a^{\rm rest}$, with intrinsic scatter $\sigma_{\rm int}=0.82\pm0.05$ dex. In the unrestricted ranking of 377 feature subsets, the classical Dainotti relation remains competitive, within the top $\sim25\%$ of the tested search spaces. The highest-ranked screened expression is $\log L_a\approx0.855\log L_{\rm peak}-0.782\log T_a^{\rm rest}$, with RMSE $\approx0.46$ and $R^2\approx0.88$. The highest-ranked runs repeatedly reduce to this prompt--plateau structure, without retaining additional spectral, absorption, or afterglow-shape variables. A direct regression gives $\log L_a=4.718+0.776\log L_{\rm peak}-0.805\log T_a^{\rm rest}$, with $R_{\rm P}=0.91$ for 246 bursts. An errors-in-variables refit gives consistent coefficients. After removing shared luminosity-distance dependence, the relation retains non-negligible explanatory power ($R^2\approx0.62$), consistent with a possible intrinsic component. Restricted Monte Carlo null tests give $p\approx0.01$. Plateau luminosity therefore shows a coupled dependence on prompt luminosity and plateau duration.}
 
\end{abstract}


\section{Introduction}

{Gamma-ray bursts (GRBs) are the most luminous transient explosions in the Universe. Their prompt emission releases a large amount of energy over a short time and is followed by X-ray, optical, and radio afterglows \citep{1999PhR...314..575P,2018pgrb.book.....Z}. The temporal structure of the afterglow provides information on the relativistic outflow and the continuing activity of the central engine \citep{1999ApJ...519L..17S,2004IJMPA..19.2385Z,2004RvMP...76.1143P, 2006RPPh...69.2259M, 2006Natur.444.1044G, 2014ARA&A..52...43B, 2015PhR...561....1K, 2015JHEAp...7...73D}.}

{A common feature of GRB X-ray afterglows is the plateau phase, during which the luminosity remains nearly constant for hundreds to thousands of seconds and then changes to a steeper power-law decay \citep{2006ApJ...642..389N, 2006ApJ...642..354Z,2009MNRAS.397.1177E}. Proposed explanations include magnetar spin-down, fallback accretion, forward-shock and density-structure models \citep{2014ApJ...789...39U}, black-hole spin-down \citep{Lenart2025}, and sustained energy injection \citep{2011A&A...526A.121D,2001ApJ...552L..35Z, Wang:2024fun,2013MNRAS.430.1061R,2015ApJ...805...89L, BERNARDINI201564, 2025ApJ...989...60Y, 2025ApJ...993...68M}.}

{An important empirical relation of the plateau phase is the Dainotti relation, which connects the plateau-end luminosity $L_a$ and the rest-frame plateau-end time $T_a^{\rm rest}$ through an approximate anticorrelation,}
\begin{equation}
\log L_a = A + B \log T_a^{\rm rest},
\end{equation}
where the slope is $B = -1.07_{-0.14}^{+0.09}$ \citep{2008MNRAS.391L..79D, 2010ApJ...722L.215D, 2011ApJ...730..135D, 2011MNRAS.418.2202D, 2013ApJ...774..157D}. 
{The Dainotti relation has been used to study plateau physics and has also been examined for GRB standardization and cosmology \citep{10.1111/j.1365-2966.2009.15456.x, 2022MNRAS.516.1386C,2022MNRAS.510.2928C, 10.1111/j.1365-2966.2010.17197.x, 2014ApJ...783..126P, Dainotti:2022ked, Favale:2024lgp,2023ApJ...951...63D,2026arXiv260318223D}.}

{The scatter of the Dainotti relation suggests that plateau luminosity may depend on additional variables. Prompt-emission properties are relevant because the prompt and plateau phases may be connected to the same central engine or energy reservoir. The three-dimensional fundamental plane includes the prompt peak luminosity $L_{\rm peak}$ and has a smaller intrinsic scatter \citep{2016ApJ...825L..20D,2017ApJ...848...88D,Dainotti:2022ked}. The scatter and normalization can also differ among GRB subclasses, including platinum, SNe-associated, and kilonova-associated samples \citep{2020ApJ...904...97D}.}

{In this work, we examine whether the Dainotti relation is the best low-dimensional description of the data or a projection of a higher-dimensional relation. Symbolic regression searches directly over mathematical expressions rather than fitting a prescribed functional form \citep[see e.g.,][]{Schmidt2010}. Using operators such as $+$, $-$, $\times$, $/$, $\exp$, $\log$, and $\sqrt{}$, it determines candidate functional forms and their constants. We use the PySR framework \citep{2023arXiv230501582C} to address four questions: whether the Dainotti relation is recovered, where it ranks among the low-dimensional candidates, whether a stronger prompt--plateau relation is present, and how robust the selected relations are.}

{We also determine how much of the selected correlation remains after removing the shared luminosity-distance scaling.}

The structure of the paper is as follows. In Section~\ref{sec:symbolic} we briefly review symbolic regression and its role in empirical relation discovery/rediscovery. In Section~\ref{sec:data} we describe the dataset including the rest-frame quantities. In Section~\ref{sec:methods} we present the symbolic search methodology, including both anchored and unrestricted searches. Section~\ref{sec:results} presents the main results, including the rediscovery and ranking of the Dainotti relation, and the prompt--plateau plane. Section~\ref{sec:null} presents the Monte Carlo null test. Section~\ref{sec:discussion} discusses the physical interpretation, limitations, and implications of the results. {A secondary redshift-prediction test is summarized in the Appendix.}

\section{{Symbolic Regression and Empirical Relation Search}}
\label{sec:symbolic}

{Empirical relations have played an important role in astrophysics. Several physical laws were first identified as regularities in observational data before their physical origin was understood. A compact analytic relation can therefore help identify an underlying pattern in a phenomenon \citep[see e.g.,][]{10.7208/chicago/9780226789583.001.0001,2012msma.book.....F,2013SADM....6...15A}.}

{Symbolic regression is a machine-learning method for this purpose. Conventional regression begins with a chosen functional form and fits its parameters. Symbolic regression instead searches over mathematical expressions and selects equations that describe the data with low complexity \citep[see e.g., ][and references therein]{2019arXiv191008892J,2022arXiv221110873M}.}

{Modern symbolic-regression algorithms construct candidate equations from specified variables and mathematical operators and optimize them through genetic programming or related evolutionary methods. \citet{Schmidt2010} showed that such methods can recover compact physical relations from suitable data. Later implementations improved their efficiency and scalability \citep{2022arXiv221110873M}.}

{Recent studies have extended symbolic regression beyond benchmark problems. In astrophysics, it has uncovered relations between dark matter and baryonic physics that simple power-law fits missed \citep{2023MNRAS.522.2628W}. It has also been used to distill black-box models: by applying symbolic regression to neural networks trained on $N$-body simulations, \cite{2020arXiv200611287C} recovered Newton's law of gravity in a compact and interpretable form. More recently, similar methods have been applied to the classification of fast radio bursts across different observational catalogs \citep{2025arXiv251204204L}. These applications demonstrate its ability to represent high-dimensional behavior with explicit equations \citep[see e.g.,][and references therein]{2023arXiv230501582C}.}

{For astrophysical applications, the method has two relevant advantages: it can scan a broad functional space with fewer assumptions about the functional form, and it returns explicit equations that can be interpreted physically. We use the PySR framework \citep{2023arXiv230501582C}, which combines evolutionary search with symbolic optimization.}

{GRB plateau emission may depend on timescales, luminosities, spectral properties, and prompt-emission observables. A symbolic search tests whether the Dainotti relation is recovered and whether other low-dimensional relations give a smaller prediction error.}

{Most analyses examine a limited number of physically motivated combinations of luminosity ($L$) and timescale ($T$). Here we compare hundreds of variable combinations with the same procedure and determine whether the Dainotti relation is preferred or represents a lower-dimensional projection of a broader correlation.}

\section{Data and Derived Quantities}
\label{sec:data}

{We analyze 251 GRBs with measured redshifts and plateau properties. The observables are the plateau flux and time, prompt peak flux, prompt fluence and duration, spectral slopes, and line-of-sight column density. The data are taken from the plateau catalog and the associated prompt and afterglow quantities used in \citet{2020ApJ...905L..26D, 2020ApJ...903...18S, 2025A&A...698A..92N}. Five bursts lack a finite prompt peak-flux measurement, so analyses requiring $L_{\rm peak}$ contain 246 objects.}

The rest-frame quantities in this catalog were derived by \citet{2025A&A...698A..92N} from the observed data using the measured redshift and assuming a flat $\Lambda$CDM cosmology with Planck 2018 parameters \citep{2020A&A...641A...6P}. Specifically, the plateau luminosity is given by $L_a = 4\pi D_L^2 F_a$, where $D_L(z)$ is the luminosity distance and $F_a$ is the observed plateau flux. The rest-frame plateau time is $T_a^{\rm rest} = T_a^{\rm obs}/(1+z)$, and the rest-frame prompt duration is $T_{90}^{\rm rest} = T_{90}^{\rm obs}/(1+z)$. The prompt peak luminosity is constructed analogously: $L_{\rm peak} = 4\pi D_L^2 F_{\rm peak}$, where $F_{\rm peak}$ is the observed prompt peak flux.

{After filtering non-finite values, the rest-frame table for the unrestricted search contains 251 bursts. We use the following 13 candidate variables, with the definitions of \citet{2025A&A...698A..92N}:}

\begin{itemize}
\item $\log T_a^{\rm rest}$: rest-frame plateau end time
\item $\log T_{90}^{\rm rest}$: rest-frame prompt duration
\item $\mathrm{PhotonIndex}$: prompt photon spectral index from Swift BAT
\item $\alpha$: temporal decay index after the plateau
\item $\beta$: spectral index of the plateau emission
\item $\Gamma$: time-averaged spectral index from Swift XRT
\item $N_H^{\rm intrinsic}$: neutral hydrogen column density along the line of sight
\item $\log F_{\rm peak}$: observed prompt peak flux
\item $\log S$: fluence of the prompt emission
\item $\log L_{\rm peak}$: prompt peak luminosity
\item $\log(S/F_{\rm peak})$: fluence-to-peak-flux ratio
\item $\log(F_{\rm peak}/T_{90})$: peak-flux-to-duration ratio
\item $\log(T_a/T_{90})$: plateau-time-to-prompt-duration ratio
\end{itemize}

{For each positive dimensional quantity, we take the base-10 logarithm of its numerical value in the cataloged units. The logarithm can therefore be negative when that numerical value is below unity. This transformation is appropriate for power-law-like astrophysical scaling relations and improves numerical stability \citep[see e.g.,][]{2005physics..11182D}. Dimensionless spectral and temporal indices are used as cataloged.}

{Both $L_a$ and $L_{\rm peak}$ contain the factor $D_L^2$, which can strengthen their correlation through cosmological scaling. In the distance-controlled analysis, we fit each luminosity-like quantity as a linear function of $\log D_L$ and use the residuals to determine how much of the relation remains after removing the shared distance dependence \citep{2012ApJ...751..149P}.}

{Although $L_{\rm iso}=4\pi D_L^2F$, a flux-limited sample can have an effective distance dependence that also reflects selection effects, population evolution, and unmodeled beaming variations. We therefore remove the empirical dependence on $\log D_L$ by residualization. This is an approximate distance control and not a model-independent correction.}

{The catalog gives measurement uncertainties for several observables. We do not pass these uncertainties to the evolutionary symbolic-regression search, because PySR optimizes a deterministic loss for each candidate expression. For the selected Dainotti and prompt--plateau forms, we include the relevant uncertainties in $\log L_a$, $\log T_a^{\rm rest}$, and $\log L_{\rm peak}$ in post-search errors-in-variables fits, following the likelihood treatment of \citet{2005physics..11182D}. We estimate coefficient stability with a non-parametric bootstrap: each of 300 resamples draws the same number of bursts with replacement, refits the selected structure, and the standard deviation of the resulting estimates is reported as the bootstrap standard error.}

{A separate 222-burst table from the same catalog release \citep{2025A&A...698A..92N} contains cataloged \texttt{Type} labels. Because many labels are composite and several subclasses contain only a few bursts, we group them into broad classes: long GRBs (LGRBs), X-ray-rich GRBs (XRRs), X-ray flashes (XRFs), supernova-associated GRBs (SNe-GRBs), ultra-long GRBs (ULGRBs), and short GRBs with extended emission (SEE). We report groups with at least 15 bursts. This class-dependent fit is used as a consistency test alongside the main 251-burst unrestricted ranking.}

\section{Methodology}
\label{sec:methods}

\subsection{Classical Dainotti Baseline and Anchored Symbolic Search}

As a baseline, we fit the classical linear Dainotti relation \citep{2008MNRAS.391L..79D}
\begin{equation}
\log L_a = A + B \log T_a^{\rm rest}
\end{equation}
using ordinary least squares \citep[OLS;][]{1990ApJ...364..104I}. {The OLS baseline is used for the validation-ranking analysis because the symbolic models are also selected by prediction loss on the same validation split. It is a predictive baseline, not an uncertainty-aware coefficient estimate.} On the validation split used throughout the main symbolic search, the classical Dainotti OLS baseline is
\begin{equation}
\label{eq:5}
\log L_a = 51.2473 - 1.1597 \log T_a^{\rm rest},
\end{equation}
with validation Pearson correlation coefficient {$R_{\rm P} = 0.7417$}, $R^2 \approx 0.535$, and $\mathrm{RMSE} = 0.9085.$

{Separately, we refit the same two-dimensional relation with the D'Agostini errors-in-variables likelihood, including uncertainties in both $\log T_a^{\rm rest}$ and $\log L_a$ and an intrinsic-scatter term \citep{2005physics..11182D}. On the 246-burst subset with complete uncertainty information, this fit gives}
\begin{equation}
{\log L_a = (51.92\pm0.30) - (1.31\pm0.09)\log T_a^{\rm rest},}
\end{equation}
{with intrinsic scatter $\sigma_{\rm int}=0.82\pm0.05$ dex. The slope is consistent with the OLS Dainotti baseline within the scatter of the sample, but the D'Agostini fit is the more appropriate estimate when the uncertainties in both variables are considered. The quoted uncertainties here and below are the non-parametric bootstrap standard errors defined above, rather than percentile confidence intervals.}

{First, we perform a Dainotti-guided search in which every candidate model contains $\log T_a^{\rm rest}$. This search tests whether symbolic regression recovers the Dainotti form and whether additional variables improve it.}

{The additional variables are drawn from the prompt, afterglow, and spectral observables, with at most three additional variables in each candidate set.}

The target variable is always $\log L_a$. Candidate expressions are constructed from basic binary operators and selected by validation performance and simplicity. {PySR optimizes the symbolic expressions with a squared-error loss; measurement uncertainties are not used in this evolutionary search. For the one-variable anchored feature set, $[\log T_a^{\rm rest}]$, the best validation-loss PySR expression is the same affine Dainotti form as the OLS validation baseline,}
\begin{equation}
{\log L_a \approx 51.247 - 1.160\log T_a^{\rm rest},}
\end{equation}
{with validation $R_{\rm P} \approx 0.742$, $R^2 \approx 0.535$, and $\mathrm{RMSE}\approx0.908.$ The match is expected: the selected one-variable PySR expression has free affine constants and is optimized with the same squared-error objective used for the predictive ranking. This expression should not be confused with the D'Agostini errors-in-variables refit above, which uses measurement uncertainties and intrinsic scatter. A lower-complexity pure symbolic one-variable expression, $\log L_a \approx 50.721-\log T_a^{\rm rest}$, also appears in the symbolic ranking and has $\mathrm{RMSE}\approx0.934$; below we use it only to track the simple symbolic Dainotti candidate in the 377-feature-subset ranking. The anchored search therefore reproduces the standard Dainotti anticorrelation.}

{We distinguish four related Dainotti quantities throughout the paper: the literature relation, the OLS baseline fitted for validation ranking, the D'Agostini uncertainty-aware refit, and the lower-complexity pure symbolic candidate used as a symbolic-ranking marker. They share the same empirical anti-correlation but need not have identical fitted intercepts, slopes, or validation losses.}

{Recovery of the Dainotti relation does not establish it as the best multivariate description. We therefore perform an unrestricted search in which $L_{\rm peak}$ and the other candidate variables are evaluated under the same procedure.}

\subsection{Unrestricted Symbolic Search}

{In the second search, $\log T_a^{\rm rest}$ is not required. We consider every non-empty subset of at most three variables from Section~\ref{sec:data}. The resulting $13+\binom{13}{2}+\binom{13}{3}=377$ feature-subset search spaces can each generate several symbolic expressions.}

{We first screen each feature subset with a shallow symbolic search and then apply a deeper search to the better-performing subsets.}

{Because $L_a$ and $L_{\rm peak}$ share the $D_L^2$ factor, we assess the selected relations using both validation metrics and distance-controlled diagnostics.}

\subsection{{Symbolic-regression validation and robustness}}

{Symbolic regression is used here as a structure-discovery tool rather than as a direct parameter-inference method. Candidate expressions are generated with PySR on the training subset and evaluated on a held-out validation subset. The main split is an unstratified random 75\%/25\% split with random seed 42; no class labels enter this split, and all model rankings are based on validation performance rather than training loss. Models are ranked primarily by validation MAE \citep[mean absolute error;][]{2005ClRes..30...79W}, which provides a robust measure of typical prediction error and is less sensitive to outliers. We also report RMSE (root-mean-square error), Pearson's $R_{\rm P}$, and $R^2$ for the final selected models. The coefficient of determination, $R^2$, quantifies the fraction of variance in $\log L_a$ explained by the model, complementing RMSE and correlation-based metrics \citep{draper1998applied}. Unless otherwise stated, all metrics are computed for the logarithmic target and refer to validation performance.}

{The same validation set is used for feature-subset screening, ranking, and final metric reporting. The reported validation metrics are therefore comparative ranking statistics, not unbiased estimates of fully independent out-of-sample performance.}

{As a post-selection stability check, we repeated the regression comparison for the selected empirical forms over 100 random 75/25 train-validation splits. This repeated-split exercise does not rerun the full symbolic-regression search and should not be interpreted as nested cross-validation for model discovery. It tests whether the comparison between the classical Dainotti relation, the prompt--plateau plane, and the literature fundamental-plane baseline depends on the random seed used for the main validation split.}

{Overfitting is controlled in several ways. The main unrestricted symbolic search is restricted to feature subsets of at most three variables, with finite expression size and depth limits in PySR. PySR was run with a parsimony penalty, and the expression complexity returned by PySR is recorded together with the validation metrics. Feature-subset search spaces are ranked primarily by validation MAE, with validation RMSE, distance-controlled MAE, and expression complexity used as secondary diagnostics. The analysis first screens all 377 feature subsets with a shallow search and then reruns the better-performing subsets with a deeper search, so the final ranking is not based on a single unconstrained symbolic-regression run. Because the symbolic expressions are discovered by an evolutionary search rather than fitted as a single pre-specified nested model family, these complexity diagnostics are used only to assess parsimony and not as formal likelihood-ratio or information-criterion model-selection tests.}

{The constants returned by symbolic regression are optimized parameters of the selected expression, not maximum-likelihood coefficients with formal covariance estimates. After symbolic regression identifies a preferred analytic structure, the corresponding variables are refit with conventional statistical methods. In particular, we use OLS for direct comparison with validation-ranked baselines and D'Agostini errors-in-variables regressions, including errors in $\log L_a$, $\log L_{\rm peak}$, and $\log T_a^{\rm rest}$, to estimate coefficient uncertainties and intrinsic scatter. The quoted uncertainties are bootstrap standard errors from 300 resamplings of the selected structure.}

{The stability of the recovered relations is checked through complementary diagnostics: comparison with established literature relations, including the classical Dainotti relation and the prompt--plateau fundamental plane; distance-controlled analyses that remove shared luminosity-distance dependence; class-dependent fits using the cataloged Type labels; and Monte Carlo null tests based on randomized target assignments. In each null realization, $\log L_a$ is randomly permuted among bursts while all predictor variables are kept fixed, preserving the one-dimensional distributions but removing burst-by-burst associations \citep{Efron1994}. These checks test whether the reported prompt--plateau relation could be an artifact of the symbolic-regression algorithm itself.}

\section{Results}
\label{sec:results}

\subsection{Unrestricted Search: Best Relations}

{Among the 377 feature subsets, the models with the lowest validation MAE converge on the same structure involving $\log L_{\rm peak}$ and $\log T_a^{\rm rest}$:}
\begin{equation}\label{eq:unres}
\log L_a \approx 0.855 \log L_{\rm peak} - 0.782 \log T_a^{\rm rest},
\end{equation}
with RMSE $=0.4589$ and $R^2 \approx 0.8815$.

{Refitting the same two predictors with the D'Agostini errors-in-variables likelihood gives}
\begin{equation}
\begin{aligned}
\log L_a ={}& (4.89\pm3.38) + (0.773\pm0.054)\log L_{\rm peak} \\
& - (0.808\pm0.076)\log T_a^{\rm rest},
\end{aligned}
\end{equation}
{with intrinsic scatter $\sigma_{\rm int}=0.512\pm0.040$ dex. These uncertainties are attached to the post-search refit of the symbolic structure, not to the evolutionary PySR constants themselves.}

{The deeper search gives simpler expressions with the same variables, typically $\log L_a\approx0.86(\log L_{\rm peak}-\log T_a^{\rm rest})$, with slightly worse performance (RMSE $=0.4748$, $R^2\approx0.8731$). This compact expression is useful for interpretation, but its two slopes are constrained to have equal magnitude. It is therefore distinct from the best screened expression in Eq.~\ref{eq:unres}. Table~\ref{tab:screened_refined} compares the two forms.}

{Table~\ref{tab:toprelations} lists the top refined symbolic relations. The feature-set column gives the variables made available to PySR in that run, while the effective-variable column shows which variables actually appear in the returned expression. Several high-ranked runs were supplied with an additional variable, such as PhotonIndex, $\Gamma$, $\beta$, $\alpha$, or $N_H^{\rm intrinsic}$, but that variable does not appear in the selected expression. PySR retains an extra feature only when it improves the validation loss enough to justify the added complexity. In these runs, once $\log L_{\rm peak}$ and $\log T_a^{\rm rest}$ are included, the additional spectral, absorption, or afterglow-shape variables are not retained. These rows should therefore not be read as distinct three-variable relations. They show that PySR returns the same two-variable prompt--plateau structure even when extra variables are available. This does not prove that the discarded variables are physically irrelevant; it shows that, within the present sample and search setup, most of the predictive information is captured by $L_{\rm peak}$ and $T_a^{\rm rest}$. The repeated recovery of the same two-variable structure across multiple supplied feature sets supports treating it as a parsimonious representation of the selected relation.}

{As an illustrative post-selection nested-model check, adding the plateau spectral index $\beta$ linearly to the OLS prompt--plateau plane produces no improvement, with $F(1,242)=0.0068$, $p=0.934$, and $\Delta R^2\simeq4.9\times10^{-6}$.}

\begin{table*}
\centering
\scriptsize
\caption{Top refined symbolic relations ranked by validation MAE. The feature-set column lists the variables supplied to PySR, while the selected-expression and effective-variable columns show what the returned expression actually uses. $R_{\rm P}$ is the validation Pearson correlation. Extra supplied variables that do not appear in the selected expression were not retained by PySR in these runs.}\label{tab:toprelations}
\resizebox{\textwidth}{!}{\begin{tabular}{clllccc}
\toprule
Rank & Feature set supplied & Selected expression & Effective variables & $R_{\rm P}$ & MAE & RMSE \\
\midrule
1 & $\log T_a^{\rm rest}$, PhotonIndex, $\log L_{\rm peak}$ & $0.8589(\log L_{\rm peak}-\log T_a^{\rm rest})$ & $\log T_a^{\rm rest}$, $\log L_{\rm peak}$ & 0.940 & 0.366520 & 0.474780 \\
2 & $\log T_a^{\rm rest}$, $\log T_{90}^{\rm rest}$, $\log L_{\rm peak}$ & $0.8589(\log L_{\rm peak}-\log T_a^{\rm rest})$ & $\log T_a^{\rm rest}$, $\log L_{\rm peak}$ & 0.940 & 0.366527 & 0.474786 \\
3 & $\log T_a^{\rm rest}$, $\Gamma$, $\log L_{\rm peak}$ & $0.8589(\log L_{\rm peak}-\log T_a^{\rm rest})$ & $\log T_a^{\rm rest}$, $\log L_{\rm peak}$ & 0.940 & 0.366531 & 0.474789 \\
4 & $\log T_a^{\rm rest}$, $\log L_{\rm peak}$, $\log(F_{\rm peak}/T_{90})$ & $0.8589(\log L_{\rm peak}-\log T_a^{\rm rest})$ & $\log T_a^{\rm rest}$, $\log L_{\rm peak}$ & 0.940 & 0.366532 & 0.474790 \\
5 & $\log T_a^{\rm rest}$, $\log L_{\rm peak}$ & $0.8589(\log L_{\rm peak}-\log T_a^{\rm rest})$ & $\log T_a^{\rm rest}$, $\log L_{\rm peak}$ & 0.940 & 0.366532 & 0.474790 \\
6 & $\log T_a^{\rm rest}$, $\log S$, $\log L_{\rm peak}$ & $0.8589(\log L_{\rm peak}-\log T_a^{\rm rest})$ & $\log T_a^{\rm rest}$, $\log L_{\rm peak}$ & 0.940 & 0.366532 & 0.474790 \\
7 & $\log T_a^{\rm rest}$, $\beta$, $\log L_{\rm peak}$ & $0.8589(\log L_{\rm peak}-\log T_a^{\rm rest})$ & $\log T_a^{\rm rest}$, $\log L_{\rm peak}$ & 0.940 & 0.366532 & 0.474790 \\
8 & $\log T_a^{\rm rest}$, $\alpha$, $\log L_{\rm peak}$ & $0.8589(\log L_{\rm peak}-\log T_a^{\rm rest})$ & $\log T_a^{\rm rest}$, $\log L_{\rm peak}$ & 0.940 & 0.366532 & 0.474790 \\
9 & $\log T_a^{\rm rest}$, $\log F_{\rm peak}$, $\log L_{\rm peak}$ & $0.8589(\log L_{\rm peak}-\log T_a^{\rm rest})$ & $\log T_a^{\rm rest}$, $\log L_{\rm peak}$ & 0.940 & 0.366533 & 0.474790 \\
10 & $\log T_a^{\rm rest}$, $N_H^{\rm intrinsic}$, $\log L_{\rm peak}$ & $0.8589(\log L_{\rm peak}-\log T_a^{\rm rest})$ & $\log T_a^{\rm rest}$, $\log L_{\rm peak}$ & 0.940 & 0.366533 & 0.474791 \\
\bottomrule
\end{tabular}
}
\end{table*}

\begin{table*}
\centering
\caption{Direct comparison of the best screened symbolic expression and the compact refined expression.}\label{tab:screened_refined}
\begin{tabular}{lccc}
\toprule
Model & Expression & MAE & RMSE \\
\midrule
Best screened symbolic & $0.8546\log L_{\rm peak}-0.7820\log T_a^{\rm rest}$ & 0.358651 & 0.458947 \\
Best refined symbolic & $0.8589(\log L_{\rm peak}-\log T_a^{\rm rest})$ & 0.366520 & 0.474780 \\
\bottomrule
\end{tabular}
\end{table*}

\begin{figure*}
\centering
\includegraphics[width=0.9\textwidth]{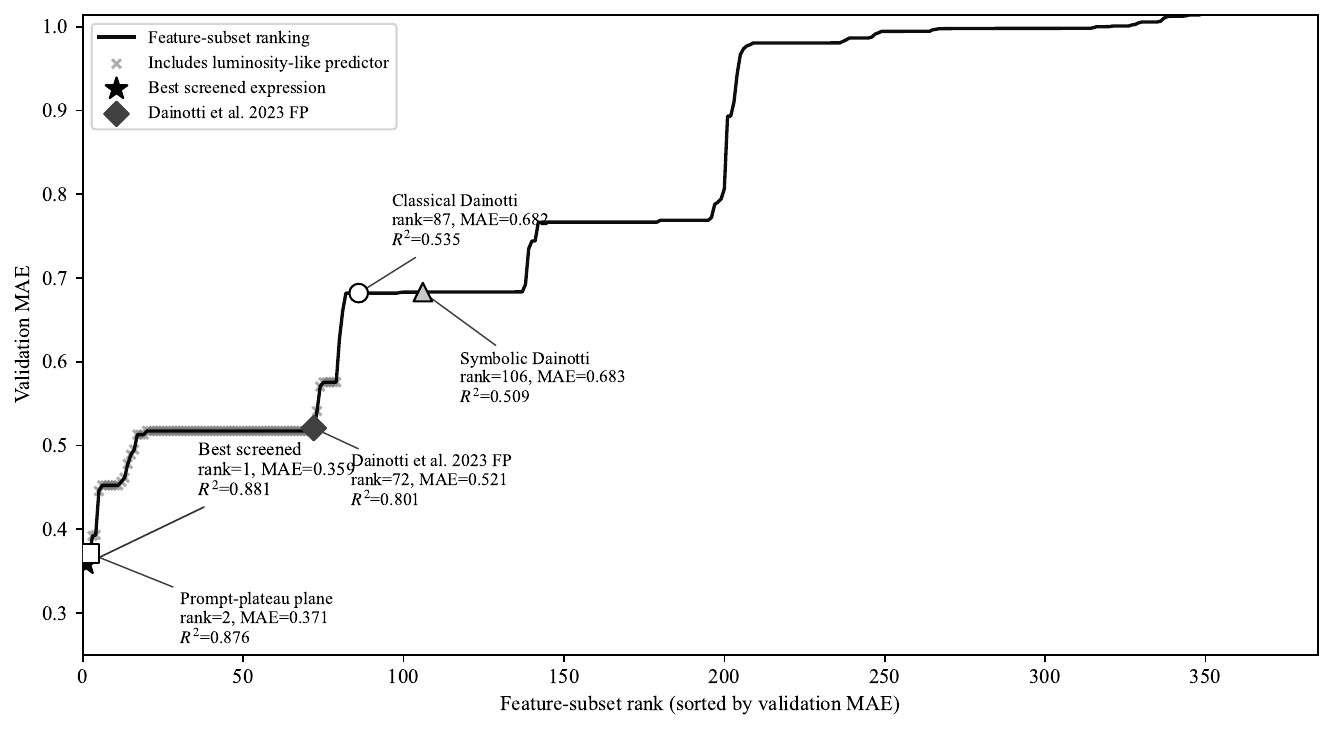}
\caption{
Ranking of all 377 unrestricted {feature-subset search spaces} explored in the search, sorted by validation MAE (lower is better). The pure symbolic Dainotti model and the classical linear Dainotti baseline are highlighted relative to the best-performing symbolic relation. {The plot also includes a literature fundamental-plane baseline using the slopes reported by \citet{Dainotti:2022ked}; only the intercept is re-normalized on the present training split because the prompt-luminosity calibration differs from that work. The Dainotti relation falls within the top $\sim25\%$ of the ranking, consistent with it being a statistically competitive correlation.} Relations that include the prompt peak luminosity $L_{\rm peak}$ {have lower validation errors}, indicating that the tightest low-dimensional {structures in this search involve} both prompt and plateau quantities.}
\label{fig:ranking377}
\end{figure*}

The pure symbolic one-variable Dainotti model, corresponding to the subset $[\log T_a^{\rm rest}]$, has validation performance {$R_{\rm P} = 0.7417$}, $\mathrm{RMSE} = 0.934,$ and $R^2 \approx 0.51.$

{These one-variable values should be distinguished from the $R_{\rm P}\simeq0.94$ values in Table~\ref{tab:toprelations}, which refer to the multivariable prompt--plateau relations containing both $\log L_{\rm peak}$ and $\log T_a^{\rm rest}$. The one-variable Dainotti relation is therefore recovered, but the higher validation correlation belongs to the three-dimensional prompt--plateau structure.}

{In the unrestricted ranking, the pure symbolic one-variable model is $106/377$ by validation MAE, placing it in the top $\sim28.0\%$ of the tested feature-subset search spaces. The classical linear Dainotti baseline performs similarly, with $R_{\rm P}=0.7417$, $\mathrm{RMSE}=0.9085$, and $R^2\approx0.53$; it ranks $87/377$, or in the top $\sim23.1\%$ (Fig.~\ref{fig:ranking377}). Both versions are therefore competitive and outperform most tested search spaces, but neither is the tightest relation in the explored parameter space.}

{The post-selection repeated-split check gives the same qualitative ordering. Over 100 random 75/25 train-validation splits, the prompt--plateau plane has a lower validation MAE than the classical OLS Dainotti relation in all 100 splits. The literature fundamental-plane baseline, evaluated with fixed published slopes and a re-normalized intercept on each training split, also outperforms the classical Dainotti relation in all 100 splits. This check does not constitute a nested validation of the symbolic-search procedure, but it shows that the main regression comparison is not tied to the particular seed-42 validation split. The results are summarized in Table~\ref{tab:repeatedsplit}.}

\begin{table*}
\centering
\caption{Post-selection repeated-split stability check for selected empirical forms. Each row summarizes 100 random 75/25 train-validation splits. The interval column gives the 16th--84th percentile range of validation MAE. The final column gives the number of splits in which the listed model has lower validation MAE than the classical OLS Dainotti relation. FP denotes the fundamental-plane baseline of \citet{Dainotti:2022ked}, evaluated with its published slopes and an intercept re-normalized on each training split. This is a stability diagnostic for selected forms, not a nested symbolic-regression cross-validation.}\label{tab:repeatedsplit}
\begin{tabular}{lcccc}
\toprule
Model & Median MAE & 16--84\% MAE & Median RMSE & Splits better than Dainotti \\
\midrule
Classical Dainotti OLS & 0.641 & 0.589--0.702 & 0.828 & --- \\
\citet{Dainotti:2022ked} FP & 0.425 & 0.387--0.477 & 0.555 & 100/100 \\
Prompt--plateau plane OLS & 0.385 & 0.351--0.433 & 0.535 & 100/100 \\
\bottomrule
\end{tabular}
\end{table*}

{The unrestricted search selects relations that include prompt luminosity. Their main structure is described by $\log L_{\rm peak}$ and $\log T_a^{\rm rest}$.}

\begin{figure*}
\centering
\includegraphics[width=0.9\textwidth]{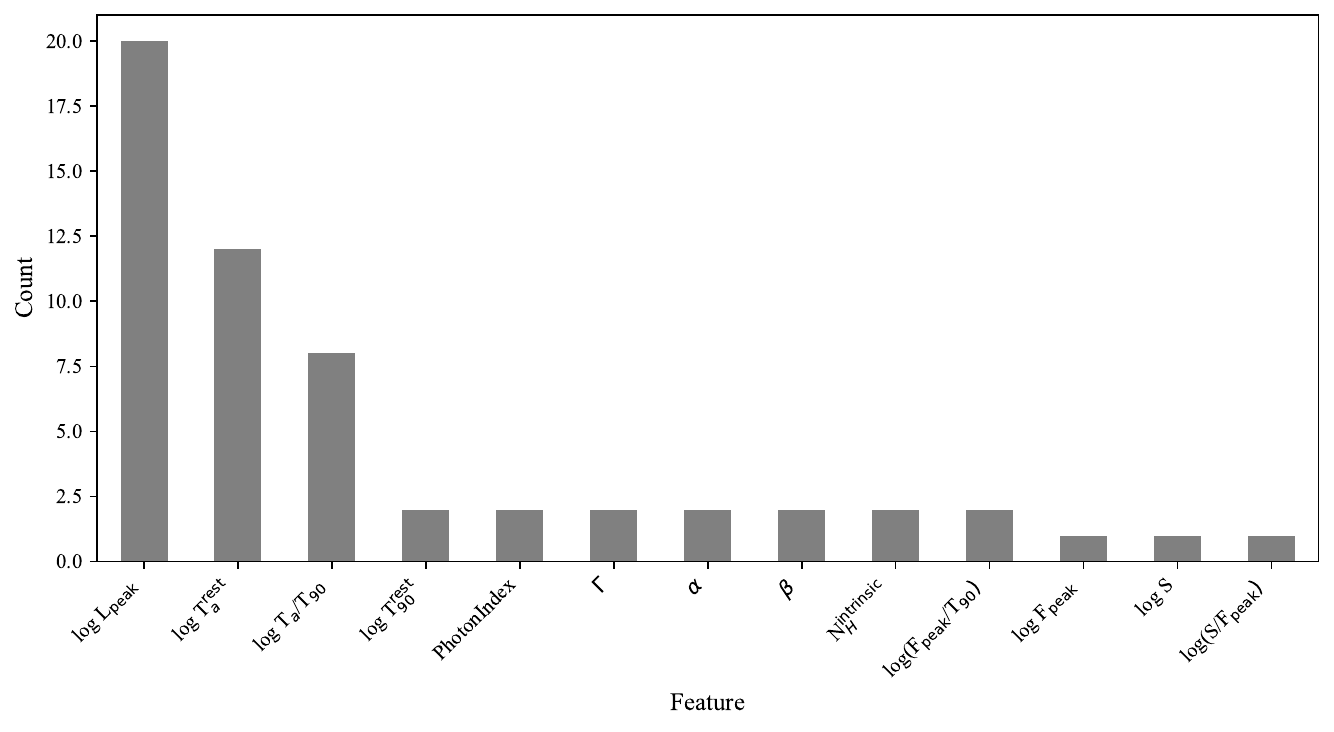}
\caption{
Feature usage among the top 20 MAE-ranked models from the unrestricted symbolic search. {This statistic refers only to the unrestricted search, where $\log T_a^{\rm rest}$ was not forced into the candidate models, and should be read as a descriptive top-20 diagnostic rather than as a formal threshold test.} Prompt peak luminosity appears in all of {these top-20} relations, suggesting that it is the variable most closely associated with the tightest plateau-luminosity correlations. By contrast, $\log T_a^{\rm rest}$ appears less often, consistent with the idea that the preferred relations involve both prompt and plateau information rather than a time-only scaling.}

\label{fig:featureusage}
\end{figure*}

{Feature-usage statistics for the top 20 screened models provide a descriptive diagnostic of the ranking. The variable $\log L_{\rm peak}$ appears in all 20 of these relations, whereas $\log T_a^{\rm rest}$ appears in 12 and $\log(T_a/T_{90})$ in 8. Other variables are much less common; see Fig.~\ref{fig:featureusage}.}

\subsection{Explicit Prompt--Plateau Plane}

{To quantify the three-dimensional prompt--plateau structure selected by symbolic regression, we fit the plane}
\[
{\log L_a = A + B \log T_a^{\rm rest} + C \log L_{\rm peak}}
\]
on the 246-burst subset with complete measurements. We obtain
\begin{equation}\label{basic:3d}
\log L_a = 4.718 + 0.776 \log L_{\rm peak} - 0.805 \log T_a^{\rm rest},
\end{equation}
with {$R_{\rm P} = 0.908$}, $\mathrm{RMSE} \approx 0.535$, and $R^2 \approx 0.824.$

{The symbolic expression and explicit plane were initially evaluated on different subsets. On the same 246-burst sample, their performance is similar: the explicit plane has a slightly lower RMSE and higher $R^2$, while the fixed symbolic expression has a slightly lower MAE. We do not assign physical significance to these small differences.}

To check that the observed correlation is not produced solely by shared luminosity-distance dependence, we carried out a distance-controlled analysis. {Specifically, for each luminosity-like quantity \(Y\), we fit \(Y=a+b\log D_L\) and define \(Y_{\rm res}=Y-(a+b\log D_L)\). The prompt--plateau plane is then refitted using the residuals of \(\log L_a\) and \(\log L_{\rm peak}\), while retaining \(\log T_a^{\rm rest}\); for symbolic predictions, the same residualization is applied to both the target and prediction before computing the diagnostic metrics.} {The same issue is often treated with partial-correlation tests in GRB plateau studies \citep{2015MNRAS.451.3898D}. After controlling for both $\log D_L$ and $\log T_a^{\rm rest}$, the partial Pearson correlation between $\log L_a$ and $\log L_{\rm peak}$ is $R_{{\rm P},{\rm partial}}\simeq0.535$ with a two-sided \(t\)-test value \(p\simeq1.8\times10^{-19}\) for 246 bursts and two control variables. The residualized distance-controlled analysis is a diagnostic robustness test, not a replacement for a full Efron--Petrosian treatment of truncation and luminosity evolution \citep{1992ApJ...399..345E,2017A&A...600A..98D}.}

\begin{figure*}
\centering
\includegraphics[width=0.9\textwidth]{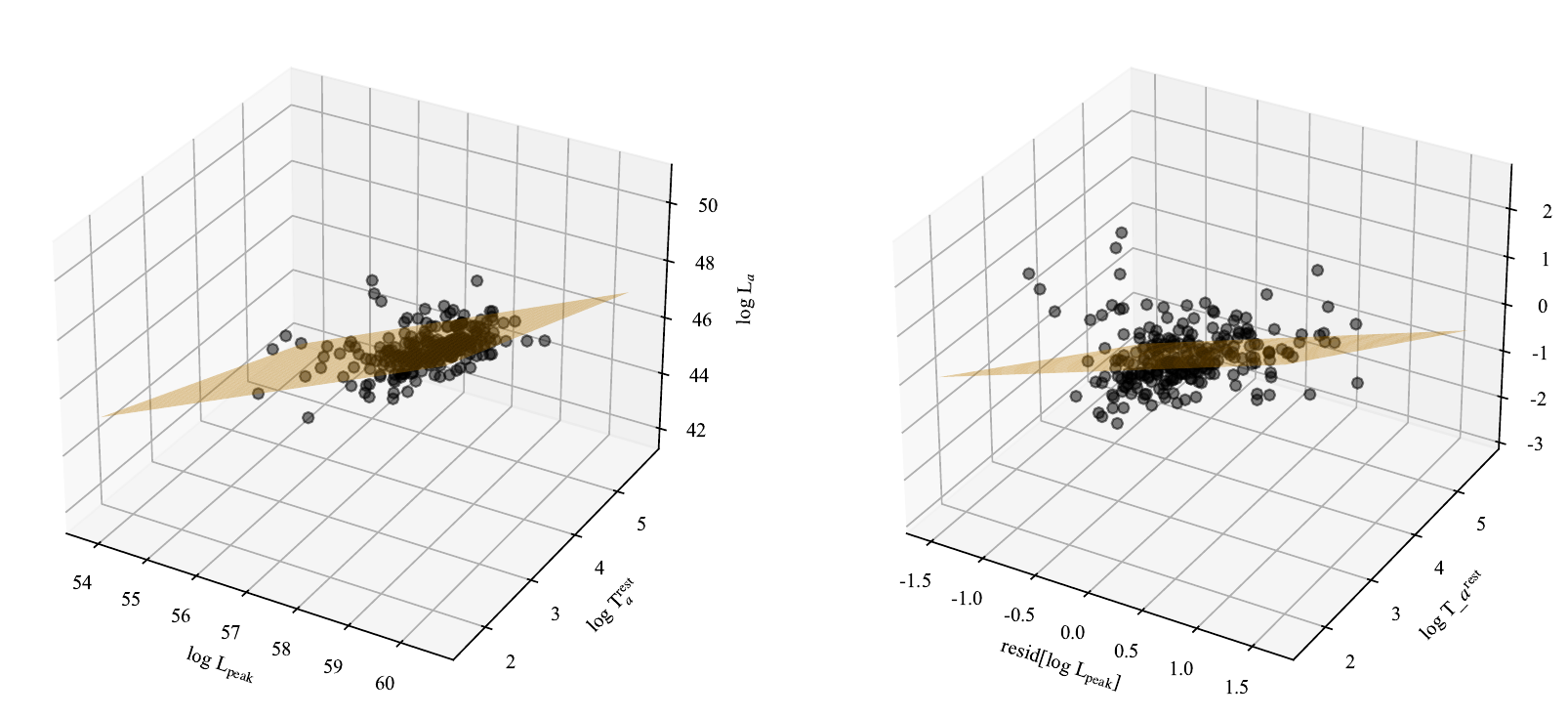}
\caption{
Raw and distance-controlled prompt--plateau planes. \textit{Left:} the plane in the space of $\log L_{\rm peak}$, $\log T_a^{\rm rest}$, and $\log L_a$, with the best-fit surface $\log L_a = 4.718 + 0.776 \log L_{\rm peak} - 0.805 \log T_a^{\rm rest}$. \textit{Right:} the same relation after removing the shared luminosity-distance dependence. {The plane remains visible in residual space, leaving room for an intrinsic component in addition to distance-driven covariance.}
}
\label{fig:plane3d}
\end{figure*}

\begin{figure*}
\centering
\includegraphics[width=0.9\textwidth]{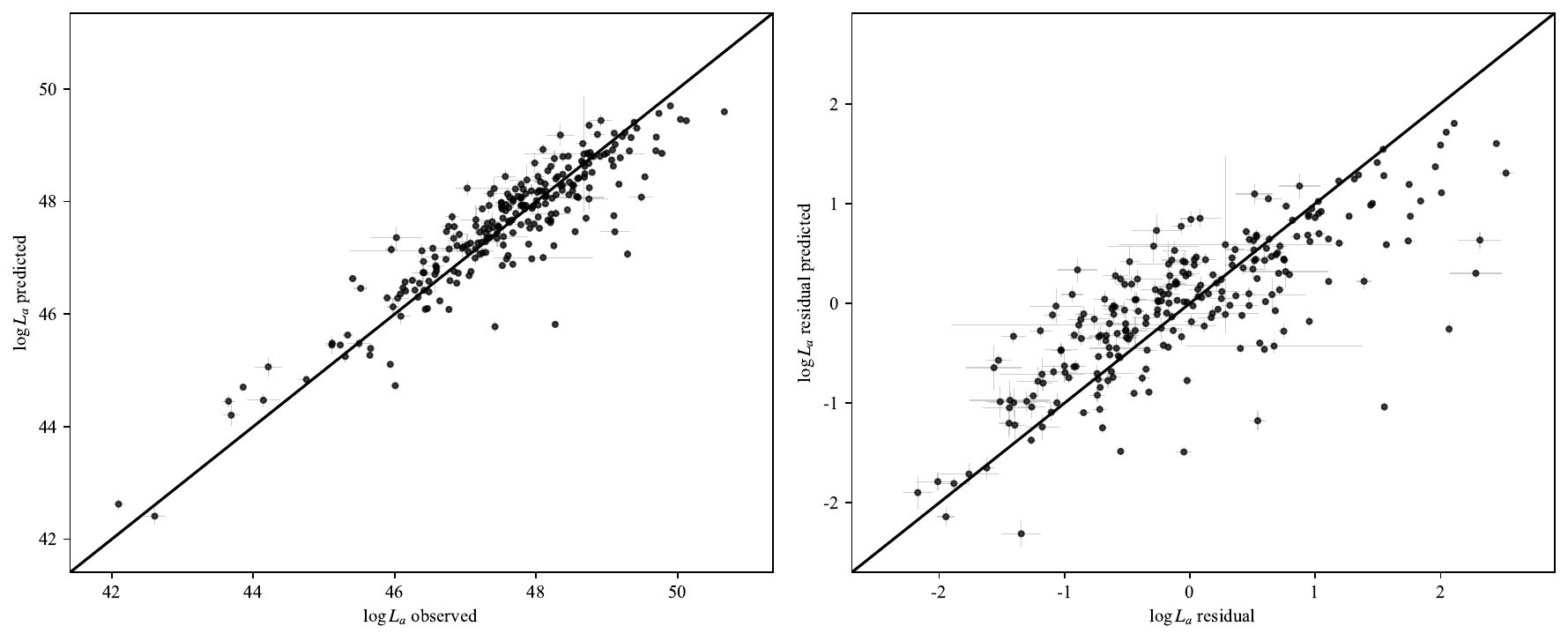}
\caption{
Predicted versus observed plateau luminosity for the explicit prompt--plateau plane fit. \textit{Left:} raw plane prediction with $\mathrm{RMSE}\approx0.535$ and $R^2\approx0.824$. \textit{Right:} distance-controlled prediction with $\mathrm{RMSE}\approx0.569$ and $R^2\approx0.622$. {The three-parameter plane remains visible after removing shared luminosity-distance scaling, although with more scatter. Horizontal bars show the catalog uncertainty in $\log L_a$, and vertical bars show the propagated uncertainty from $\log L_{\rm peak}$ and $\log T_a^{\rm rest}$ through the fitted plane.}}

\label{fig:planepred}
\end{figure*}

{After removing the shared luminosity-distance dependence, the relation retains non-negligible fit quality ($\mathrm{RMSE}\approx0.569$, $R^2\approx0.62$), leaving room for an intrinsic component in addition to cosmological scaling.} The raw and distance-controlled prompt--plateau planes, together with the corresponding predicted-versus-observed comparisons, are shown in Figs.~\ref{fig:plane3d} and \ref{fig:planepred}.

{The persistence of the relation after residualization argues against it being set entirely by shared distance scaling, although cosmological evolution and selection effects still require a dedicated truncation analysis.}

\begin{figure*}
\centering
\includegraphics[width=0.9\textwidth]{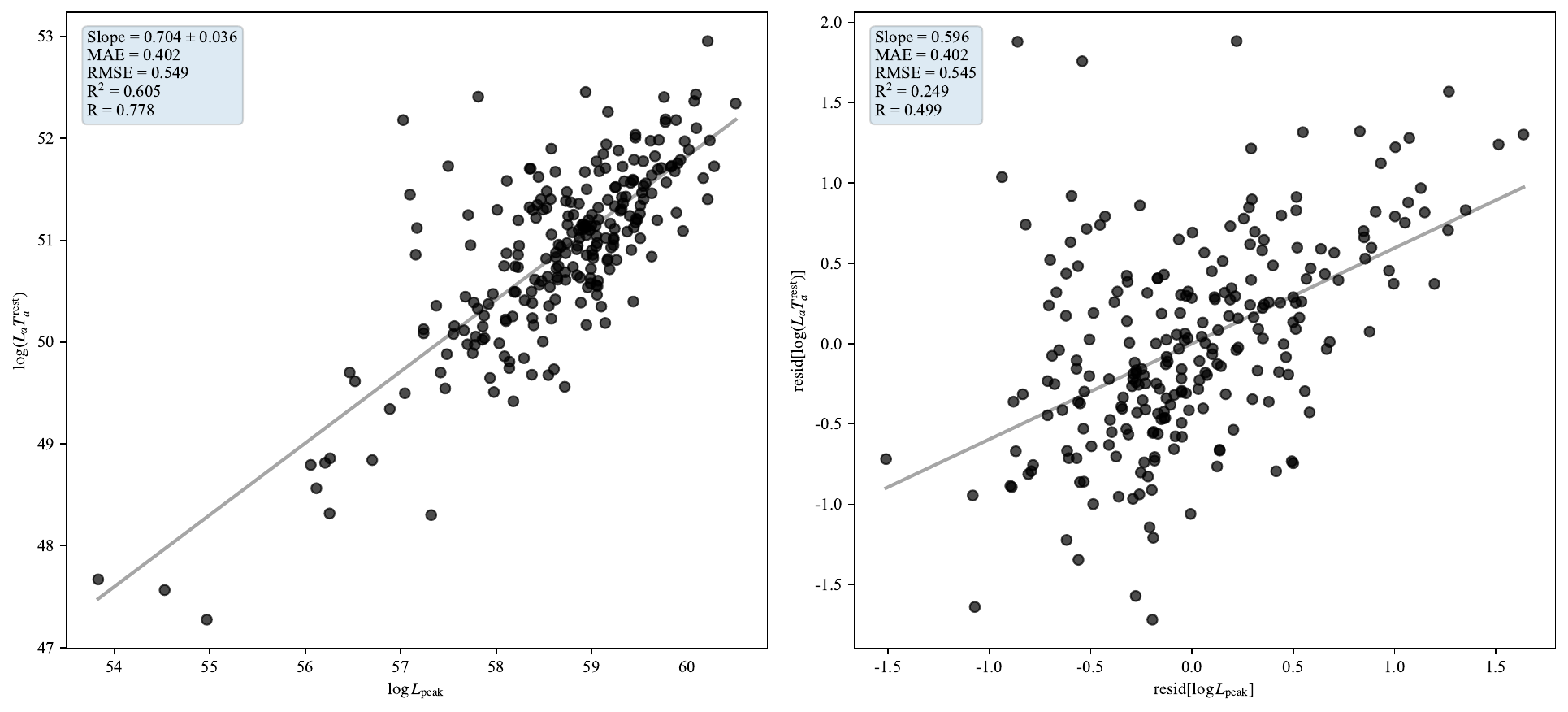}
\caption{
Plateau energy proxy, $E_{\rm plateau}=L_aT_a^{\rm rest}$, versus prompt peak luminosity. \textit{Left:} raw relation showing {a positive OLS correlation with slope $0.704\pm0.036$}, {$R_{\rm P}=0.778$}, and $R^2\approx0.605$. \textit{Right:} distance-controlled relation after removing the shared luminosity-distance dependence. The correlation weakens substantially ({$R_{\rm P}=0.499$}, $R^2\approx0.249$), which indicates that part of the raw trend comes from cosmological scaling. The remaining trend {is consistent with a weaker prompt--plateau energetics link}.}

\label{fig:eplateau}
\end{figure*}

\subsection{Energy-Based Diagnostics}

{The selected relation also connects the prompt and plateau energetics. We define the plateau energy proxy as $E_{\rm plateau}\equiv L_aT_a^{\rm rest}$. From Eq.~\ref{basic:3d}, it scales as $E_{\rm plateau}\propto L_{\rm peak}^{0.776}T_a^{0.195}$.}

{Energy-based plateau diagnostics have been discussed previously, in particular in the context of the luminosity--time and luminosity--luminosity correlations tabulated by \citet{2011MNRAS.418.2202D}. Here the same quantity is used to rewrite the prompt--plateau plane as an energy proxy and to test the limiting form $L_a\propto L_{\rm peak}/T_a^{\rm rest}$.}

{In the observed frame, plateau energy is positively correlated with prompt luminosity. The correlation becomes weaker after removing the common luminosity-distance dependence, showing that cosmological scaling contributes to the raw trend.}

{An ordinary-least-squares fit to the diagnostic relation between plateau energy and prompt luminosity yields}
\begin{equation}
\begin{aligned}
\log(E_{\rm plateau}) ={}& (9.584\pm2.136) \\
& +(0.704\pm0.036)\log L_{\rm peak},
\end{aligned}
\end{equation}
{where the uncertainties are the conventional OLS standard errors. The fit has $R_{\rm P}=0.778$ and $R^2\approx0.61$. The large intercept uncertainty results from extrapolation to $\log L_{\rm peak}=0$, which is far outside the observed range. The slope supports a prompt--plateau connection, while the distance-controlled result shows that the common luminosity-distance term strengthens the raw correlation. Figure~\ref{fig:eplateau} shows both cases.}

{After removing the luminosity-distance dependence, the relation decreases to $R^2\approx0.25$. Thus, a substantial part of the raw correlation is associated with the shared cosmological scaling.}

{Compared with the plateau-correlation tables of \citet{2011MNRAS.418.2202D}, the main difference here is that the prompt luminosity enters explicitly. The positive but sublinear slope, $\log E_{\rm plateau}\propto(0.704)\log L_{\rm peak}$ in the raw fit and a weaker residual trend after distance control, is consistent with an energy-budget connection between prompt and plateau phases, but it is not equivalent to a constant plateau-energy reservoir.}

{We also examine the ratio $L_aT_a^{\rm rest}/L_{\rm peak}$, which is constant for the limiting relation $L_a\propto L_{\rm peak}/T_a$. The observed scatter is $\sigma_{\log(L_aT_a/L_{\rm peak})}=0.619$, and its Pearson correlation with $\log T_a^{\rm rest}$ is $R_{\rm P}=0.401$.}

\begin{figure}
\centering
\includegraphics[width=\columnwidth]{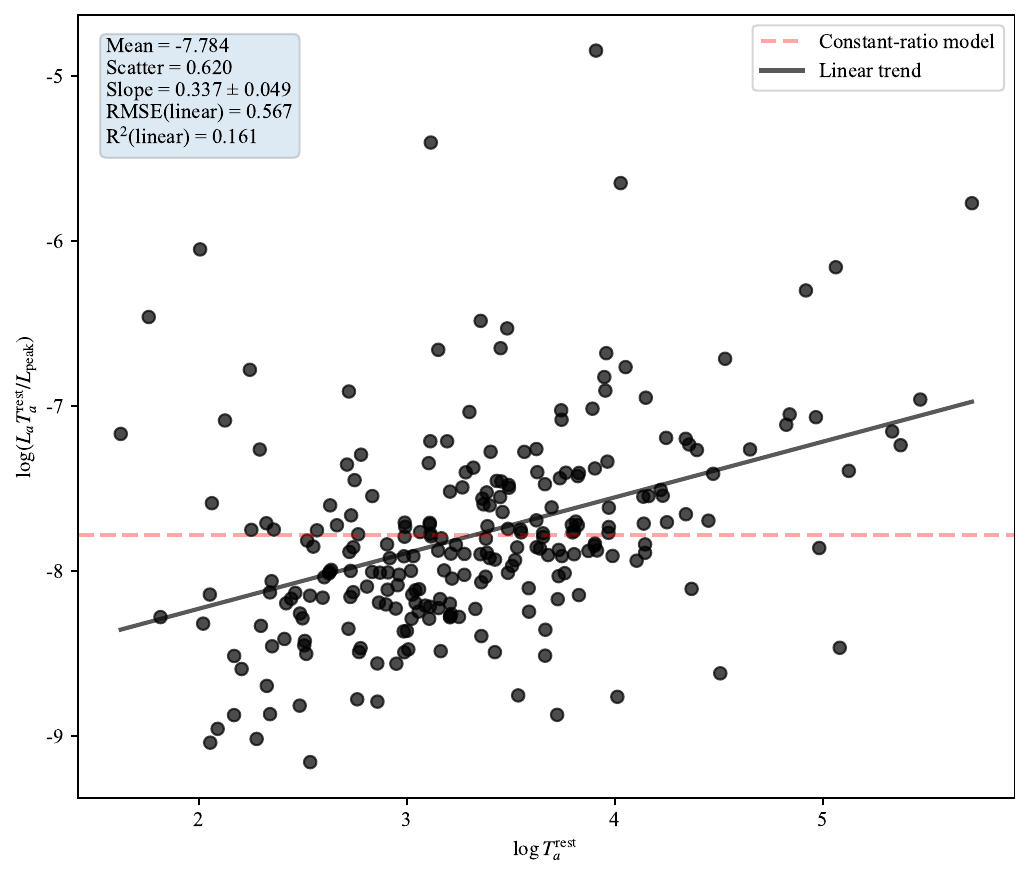}
\caption{
Diagnostic test of the symbolic relation using the ratio $\log(L_a T_a^{\rm rest}/L_{\rm peak})$ as a function of $\log T_a^{\rm rest}$. A constant-ratio model (horizontal dashed line) would correspond to $L_a \propto L_{\rm peak}/T_a$. The data instead show substantial scatter ($\sigma \approx 0.62$) and {an OLS residual trend with slope $0.337\pm0.049$} ({$R_{\rm P}\approx0.40$}, $R^2\approx0.16$). The ratio is therefore not constant, {supporting a prompt--plateau description} beyond a one-dimensional constant-ratio scaling.}

\label{fig:ratiotest}
\end{figure}

{An OLS fit gives a slope of $0.337\pm0.049$, where the uncertainty is the conventional OLS standard error. The ratio is therefore not constant and increases with duration. Figure~\ref{fig:ratiotest} shows this diagnostic. Because the common $D_L^2$ factor cancels in $L_aT_a^{\rm rest}/L_{\rm peak}$, the ratio provides an approximate distance-independent test of the constant-ratio form.}

\subsection{{Class-dependent coefficient check}}

{The main 251-burst working table does not include class labels, but the accompanying 222-burst table does. Following the class-segregation concern emphasized by \citet{2020ApJ...904...97D}, we use that reduced catalog to refit the prompt--plateau structure within broad classes defined from the composite \texttt{Type} labels. A full 377-subset PySR search is not stable for most individual classes because several contain too few bursts. We therefore re-evaluate the coefficients of the symbolic-regression-selected structure, $\log L_a \sim \log L_{\rm peak}+\log T_a^{\rm rest}$, by ordinary least squares separately in each class with at least 15 bursts. These exploratory within-class coefficients, with bootstrap standard errors, are reported as a sign-and-stability check rather than as definitive class-calibrated relations.}

{The resulting class-dependent coefficients are given in Table~\ref{tab:classcoeff}. In the two largest groups, XRR ($N=93$) and LGRB ($N=72$), the fitted coefficients retain the same signs as the full-sample plane: a positive dependence on $\log L_{\rm peak}$ and a negative dependence on $\log T_a^{\rm rest}$. The smaller SNe-GRB ($N=27$) and short/SEE-like ($N=18$) groups show the same qualitative sign pattern, but their sample sizes are too small for a stable subclass-specific symbolic-regression search. We treat this analysis as a consistency check, not as a separate class-calibrated relation.}

\begin{table}
\centering
\caption{{Class-dependent OLS prompt--plateau coefficients. Uncertainties are non-parametric bootstrap standard errors from 300 within-class resamples with replacement.}}\label{tab:classcoeff}
\resizebox{\columnwidth}{!}{\begin{tabular}{lccccc}
\toprule
{Class} & {$N$} & {$\log L_{\rm peak}$ slope} & {$\log T_a^{\rm rest}$ slope} & {RMSE} & {$R^2$} \\
\midrule
{XRR} & {93} & {$0.599\pm0.117$} & {$-0.842\pm0.120$} & {0.455} & {0.780} \\
{LGRB} & {72} & {$0.770\pm0.088$} & {$-0.907\pm0.099$} & {0.448} & {0.798} \\
{SNe-GRB} & {27} & {$0.816\pm0.133$} & {$-0.782\pm0.217$} & {0.644} & {0.864} \\
{Short/SEE} & {18} & {$0.852\pm0.123$} & {$-0.466\pm0.170$} & {0.476} & {0.709} \\
\bottomrule
\end{tabular}
}
\end{table}

\subsection{Comparison with the Three-Dimensional Dainotti Relation}

{Previous three-dimensional extensions of the Dainotti relation include prompt-emission luminosity and obtain a smaller scatter in plateau luminosity. These relations are commonly written as}

\begin{equation}
\log L_a = A + B \log T_a^{\rm rest} + C \log L_{\rm peak}.
\end{equation}

Typical coefficients reported in the literature are $B \approx -0.9$ and $C \sim 0.5$--$0.6$ \citep{2016ApJ...825L..20D,2017ApJ...848...88D,Dainotti:2022ked}, again indicating that plateau luminosity depends on both plateau duration and prompt peak luminosity. {For example, \citet{Dainotti:2022ked} report an unevolved fundamental plane with slopes $B=-0.86\pm0.13$ and $C=0.56\pm0.12$, and an evolution-corrected version with $B=-0.85\pm0.12$ and $C=0.49\pm0.13$. In the same notation, our explicit regression plane in Eq.~\ref{basic:3d} has $B=-0.805$ for $\log T_a^{\rm rest}$ and $C=0.776$ for $\log L_{\rm peak}$, with $R_{\rm P} = 0.908$, $\mathrm{RMSE} \approx 0.535$, and $R^2 \approx 0.824$. After removing the shared luminosity-distance dependence, the relation remains evident, with $\mathrm{RMSE} \approx 0.569$ and $R^2 \approx 0.62$, which is consistent with a possible intrinsic component.}

{Using the \citet{Dainotti:2022ked} slopes as a fixed-slope baseline on our validation split, we re-normalize only the intercept on the training set as \(\hat A=\langle\log L_a-C\log L_{\rm peak}-B\log T_a^{\rm rest}\rangle_{\rm train}\). This accounts for the different Swift-BAT prompt-luminosity calibration and gives validation MAE $\simeq0.521$, RMSE $\simeq0.659$, and $R^2\simeq0.80$. This is worse than the best symbolic prompt--plateau relation, but still much better than the two-dimensional Dainotti baseline.}

{In scaling form, the explicit regression gives $L_a \propto L_{\rm peak}^{0.78} T_a^{-0.81}$, which has the same structure as the symbolic forms.}

{The differences between coefficients can result from observational effects, sample selection, and fitting method. Equation~\ref{eq:unres} gives the best screened expression, $\log L_a\approx0.855\log L_{\rm peak}-0.782\log T_a^{\rm rest}$, while the deeper search gives the simpler approximation $\log L_a\approx0.86(\log L_{\rm peak}-\log T_a^{\rm rest})$.}

{Both the symbolic and regression analyses select the same prompt--plateau variables. In this description, the classical Dainotti relation is a one-dimensional projection of the broader relation.}

\begin{table}
\centering
\caption{Comparison of plateau luminosity relations.}\label{tab:tab1}
\resizebox{\columnwidth}{!}{\begin{tabular}{lcc}
\toprule
Relation & $\log T_a^{\rm rest}$ slope & $\log L_{\rm peak}$ slope \\
\midrule
Classical Dainotti (this work) & $-1.16$ & --- \\
3D Dainotti (literature) & $\sim -0.9$ & $\sim 0.5$--$0.6$ \\
{\citet{Dainotti:2022ked} slopes} & {$-0.86\pm0.13$} & {$0.56\pm0.12$} \\
Plane fit (this work) & $-0.805$ & $0.776$ \\
{D'Agostini plane (this work)} & {$-0.808\pm0.076$} & {$0.773\pm0.054$} \\
Symbolic (screened) & $-0.782$ & $0.855$ \\
Symbolic (simpler form) & $-0.86$ & $0.86$ \\
\bottomrule
\end{tabular}
}
\end{table}

\subsection{Test of Nonlinearities}

{Unary operators are single-argument functions such as $\exp(x)$, $\log(x)$, $\sqrt{x}$, and $x^2$. We include these operators in the deep-refinement stage to test whether nonlinear functions improve the selected relation.}

{The expanded search does not improve on the linear combination $\log L_{\rm peak}-\log T_a^{\rm rest}$. Expressions containing $\sqrt{\cdot}$ or $\log(\cdot)$ are either equivalent to simpler forms, such as $\exp(\log x)=x$, or have larger validation errors. The unary operators therefore provide no improvement in this search.}

{Within the present validation ranking, the empirical trend is adequately described by a power-law scaling in linear space, $L_a \propto L_{\mathrm{peak}}^{\alpha} T_a^{-\alpha},$ with $\alpha \sim 0.7$--0.9.}

{We therefore use the simpler and more interpretable scaling for this sample.}
\section{Monte Carlo Null Test}
\label{sec:null}

{We use 100 Monte Carlo realizations to test whether the symbolic relation can arise from random associations. In each realization, $\log L_a$ is permuted among the bursts while the predictors remain unchanged. This preserves the one-dimensional distributions and removes the burst-by-burst associations.}

{For each realization, we repeat a fast symbolic search over the five highest-ranked feature sets in Table~\ref{tab:toprelations} and record the validation and distance-controlled metrics. This restricted null test does not repeat the complete 377-feature-subset analysis. The null distributions are then compared with the real-data result.}

For validation performance, the best null models have typical values
\begin{equation}
\langle \mathrm{MAE}_{\rm null} \rangle \approx 0.9,
\qquad
\langle \mathrm{RMSE}_{\rm null} \rangle \approx 1.21,
\end{equation}
while the {refined real-data relation used as the benchmark in this restricted null test} achieves
\begin{equation}
\begin{aligned}
\mathrm{MAE}_{\rm real} &\approx 0.367,\\
\mathrm{RMSE}_{\rm real} &\approx 0.475,\\
R^2_{\rm real} & \approx 0.873.
\end{aligned}
\end{equation}

Similarly, for the distance-controlled metrics, the null realizations cluster around
\begin{equation}
\langle \mathrm{MAE}_{\rm null}^{\rm DC} \rangle \approx 0.90,
\qquad
\langle \mathrm{RMSE}_{\rm null}^{\rm DC} \rangle \approx 1.19,
\end{equation}
whereas the real-data relation yields
\begin{equation}
\mathrm{MAE}_{\rm real}^{\rm DC} \approx 0.368,
\qquad
\mathrm{RMSE}_{\rm real}^{\rm DC} \approx 0.469.
\end{equation}

{In all cases, no null realization matched or exceeded the performance of the real-data model. Given the finite number of realizations, we use the add-one correction \(p_{\rm corr}=(m+1)/(N_{\rm null}+1)\), where \(m\) is the number of null realizations at least as extreme as the real result. Here \(m=0\) and \(N_{\rm null}=100\), giving $p_{\rm corr}=1/101\approx0.0099$.}

\begin{figure*}
\centering
\includegraphics[width=0.49\textwidth]{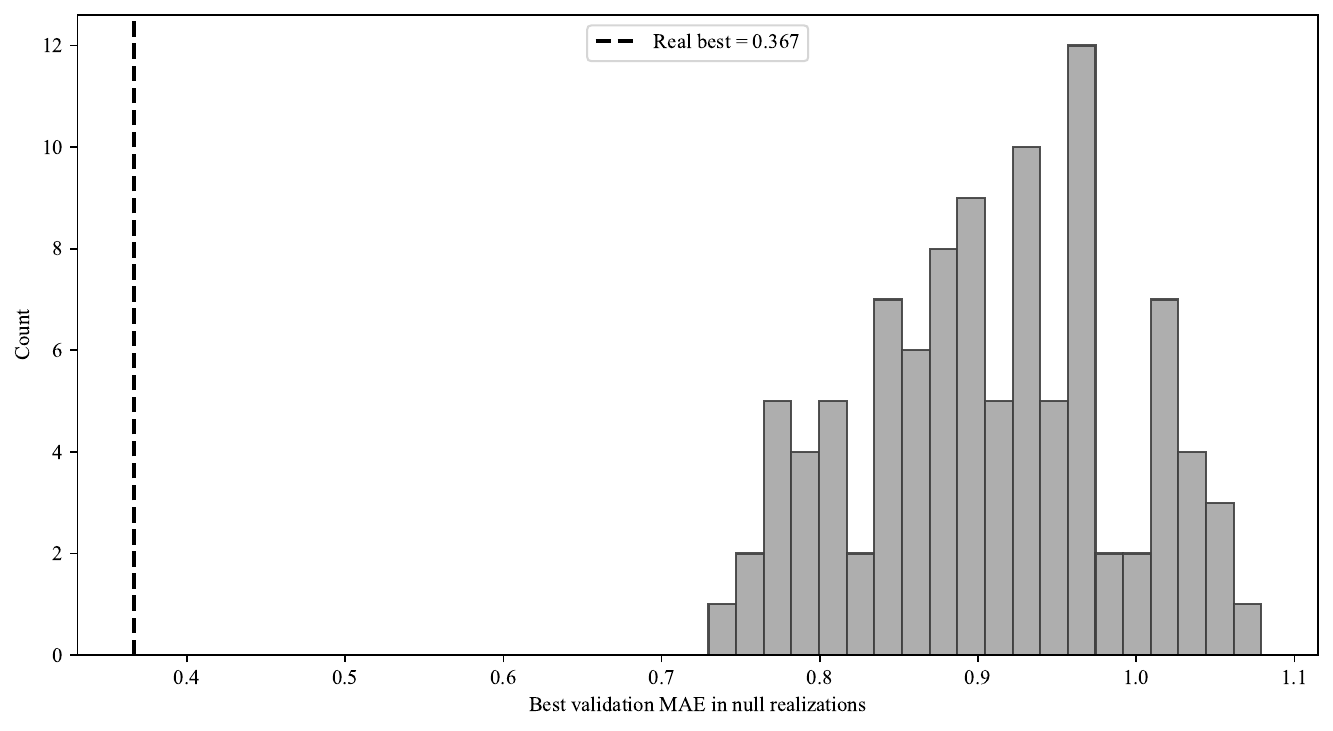}
\includegraphics[width=0.49\textwidth]{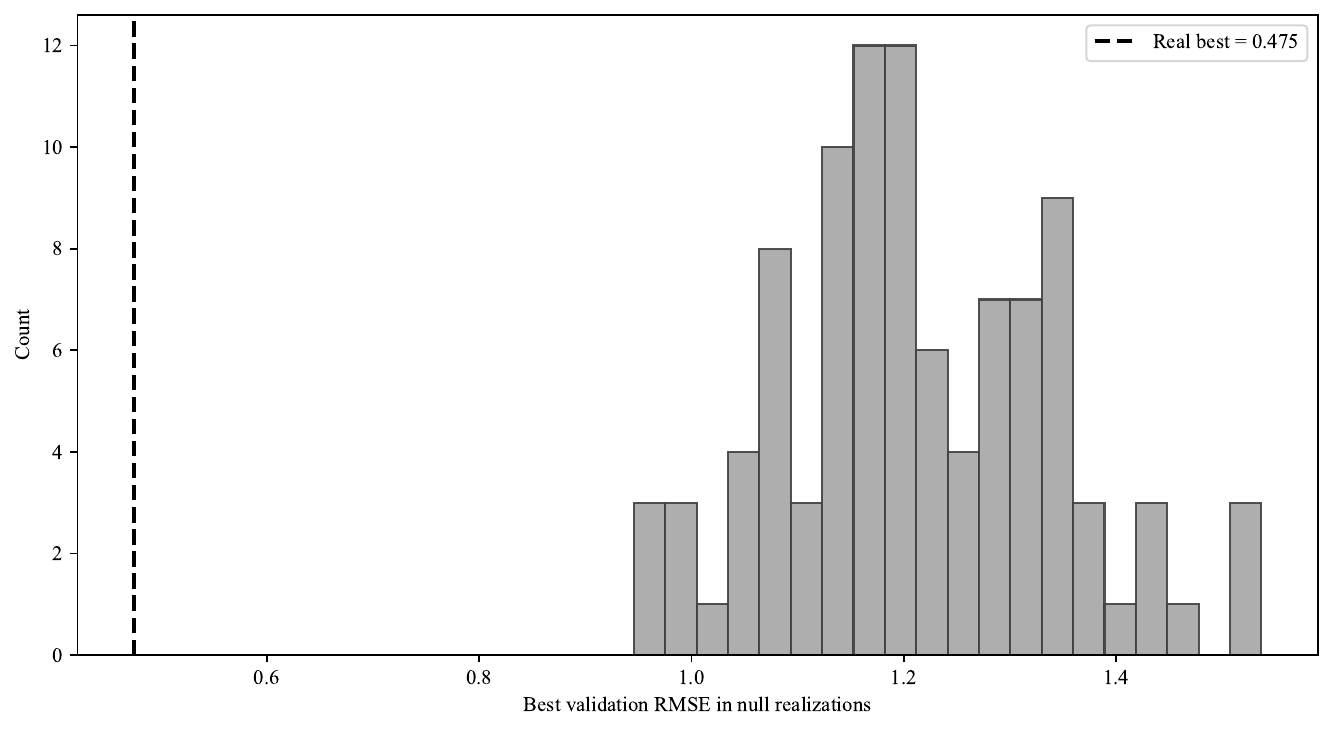}
\caption{
Monte Carlo null-test distributions for validation performance. \textit{Left:} best validation MAE from shuffled datasets. \textit{Right:} best validation RMSE. In both panels, the vertical dashed line marks the {refined real-data relation used for this comparison} (MAE $\approx 0.367$, RMSE $\approx 0.475$), which lies far outside the restricted null distributions. The null models cluster at much worse values (MAE $\sim 0.9$, RMSE $\sim 1.2$), indicating that the recovered relation is not reproduced by these random permutations of the data.}

\label{fig:null_validation}
\end{figure*}

\begin{figure*}
\centering
\includegraphics[width=0.49\textwidth]{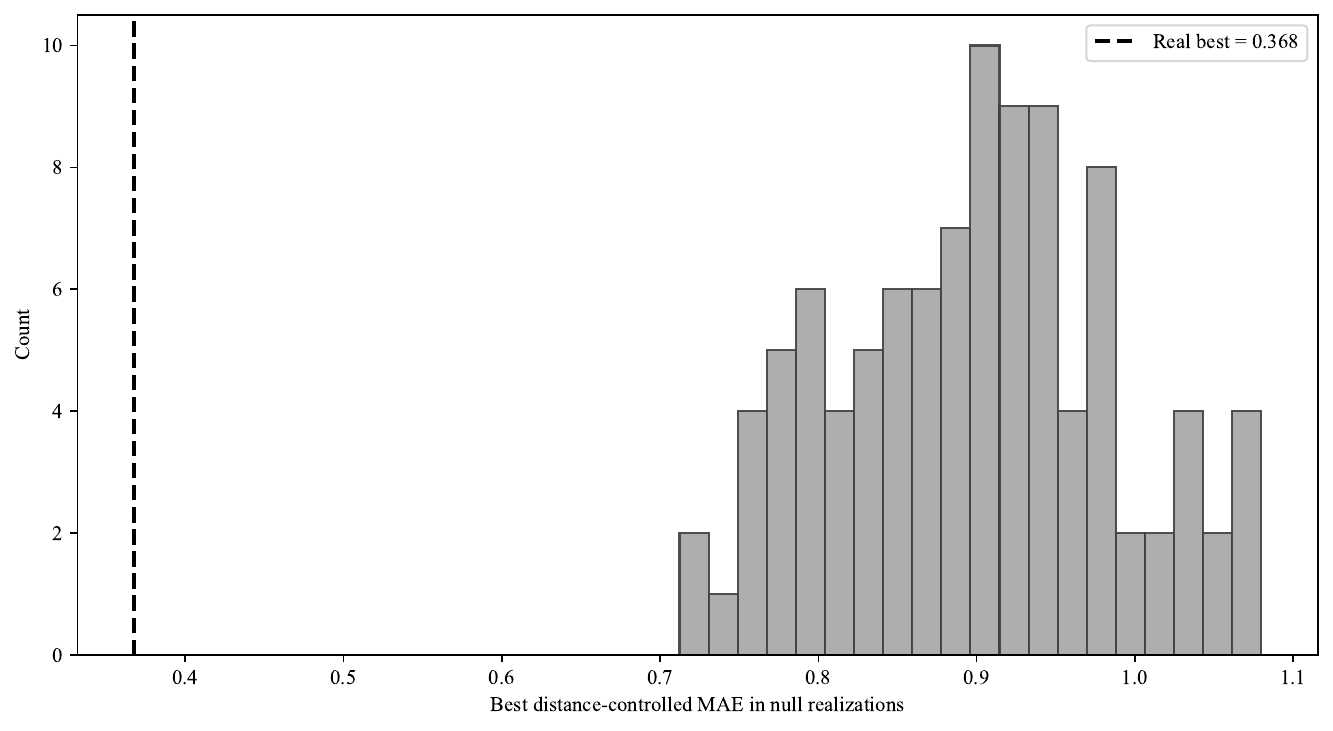}
\includegraphics[width=0.49\textwidth]{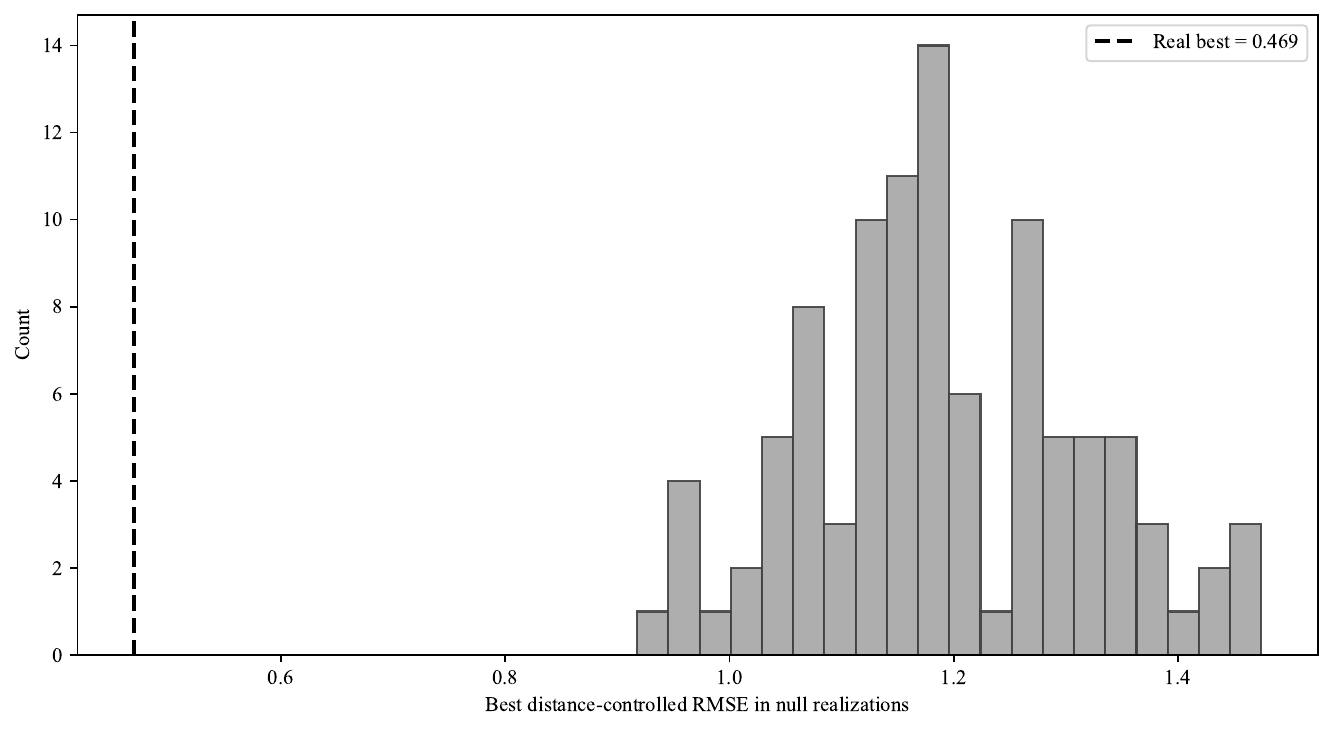}
\caption{
Monte Carlo null-test distributions for distance-controlled performance. \textit{Left:} best distance-controlled MAE. \textit{Right:} best distance-controlled RMSE. The {refined real-data relation used for this comparison} (MAE $\approx 0.368$, RMSE $\approx 0.469$) again lies well away from the restricted null distributions, showing that the real-data performance remains separated from the null results in this diagnostic.}

\label{fig:null_dc}
\end{figure*}

{As shown in Fig.~\ref{fig:null_validation} and Fig.~\ref{fig:null_dc}, the refined real-data relation lies outside the restricted null distributions in both validation space and the distance-controlled diagnostic. The null test therefore does not reproduce the observed prompt--plateau association under randomized target assignments.}

{The randomized realizations do not reproduce the performance of the symbolic relation. Because the strongest real-data models contain $L_{\rm peak}$, this result supports a statistically significant prompt--plateau association within the tested variables and restricted null design. It does not establish a new relation independently of the adopted search space.}

{The relation remains separated from the restricted null results after the distance control, although the shared luminosity-distance scaling contributes to the observed correlation. Figures~\ref{fig:null_validation} and \ref{fig:null_dc} show the validation and distance-controlled null distributions.}

\section{Discussion}
\label{sec:discussion}

{The Dainotti relation reappears in the anchored symbolic search and remains competitive in the unrestricted search. Its recovery across these analyses argues against a numerical artifact of a single search configuration, although its position in the unrestricted ranking shows that it is not the statistically tightest relation available.}

{The unrestricted search selects a higher-dimensional relation in which plateau luminosity depends on prompt luminosity and plateau duration. Equation~\ref{basic:3d} gives $\log L_a=4.718+0.776\log L_{\rm peak}-0.805\log T_a^{\rm rest}$. In linear space, $L_a\propto L_{\rm peak}^{0.78}T_a^{-0.81}$, with sublinear dependence on both variables.}

{Both exponents are close to unity, which is physically suggestive, but this should not be overinterpreted as a fixed energy-conversion law. If plateau emission were a constant-efficiency release of a prompt-energy proxy over the plateau timescale, the quantity $L_aT_a^{\rm rest}/L_{\rm peak}$ would be approximately constant. Instead, the fitted scaling gives $L_aT_a^{\rm rest}/L_{\rm peak}\propto L_{\rm peak}^{-0.22}(T_a^{\rm rest})^{0.19}$, and the observed ratio shows both substantial scatter and a residual trend with $T_a^{\rm rest}$. This behavior is consistent with event-to-event variation in the prompt--plateau energy scaling, not with a universal conversion efficiency.}

{In this diagnostic, plateau energy is not constant and increases with prompt luminosity. A luminosity--time relation alone does not include this dependence on prompt emission. Possible interpretations include long-lived magnetar spin-down \citep{2001ApJ...552L..35Z}, fallback-regulated accretion \citep{2008MNRAS.388.1729K}, and structured energy injection \citep{1998A&A...333L..87D, 2006ApJ...642..354Z}. In these models, the prompt and plateau emissions can represent different phases or channels of the same central engine.}

{The classical Dainotti relation can therefore be treated as a projection of the prompt--plateau plane, consistent with previous three-dimensional studies \citep{2016ApJ...825L..20D,2017ApJ...848...88D,Dainotti:2022ked}. Earlier analyses usually give a weaker dependence on $L_{\rm peak}$, with $C\sim0.5$--$0.6$ in the notation of Section~\ref{sec:results}, whereas both our direct and symbolic fits give a slope near $0.8$. This difference can result from fitting method, sample selection, and observational biases, and we do not assign it a separate physical interpretation. The relevant result here is that the search repeatedly selects $L_{\rm peak}$ and $T_a^{\rm rest}$ and ranks their combination above the classical Dainotti relation.}

{The feature-usage diagnostic gives the same qualitative message: $\log L_{\rm peak}$ appears in all of the top-20 screened models, whereas $\log T_a^{\rm rest}$ appears in only a subset of them. Prompt energetics carry especially strong predictive information for the plateau luminosity in this sample, while the plateau duration remains essential for the most interpretable prompt--plateau relation and for connecting the result to the classical Dainotti scaling.}

{Symbolic regression therefore recovers the known correlation and identifies the variables that occur repeatedly in the better-performing relations.}

{The prompt--plateau plane could, in principle, be obtained with traditional regression, but that approach requires the analyst to guess the relevant variables and functional form in advance. The symbolic search instead evaluates 377 feature-subset search spaces under one consistent procedure. Its recovery of the same basic plane identified in more hypothesis-driven work lends support to both approaches, while the ranking analysis adds quantitative context for the classical Dainotti relation. This is particularly useful in multivariate GRB problems, where the physically relevant combinations of observables are not always obvious beforehand.}

\subsection{{Shared Luminosity-Distance Scaling}}

{Both $L_a$ and $L_{\rm peak}$ contain the factor $D_L^2$, so cosmological scaling can strengthen their correlation. After residualizing the luminosity-like quantities against $\log D_L$, the prompt--plateau relation retains non-negligible fit quality. This is consistent with a possible intrinsic component, but the procedure does not correct for sample truncation or luminosity evolution.}

{We therefore treat the symbolic relation as a rest-frame phenomenological correlation whose interpretation must include this covariance. The main result is the ranking of the Dainotti relation among 377 feature-subset search spaces and the selection of a tighter prompt--plateau plane. We interpret this plane as a preferred low-dimensional relation among the available observables, not as a new fundamental law.}

{A full Efron--Petrosian analysis with well-defined survey truncation boundaries is needed to separate luminosity evolution from selection bias. We therefore do not interpret the residualized correlation as a selection-corrected intrinsic relation.}

\subsection{Limitations and Future Work}

{The sample is substantial for plateau studies but remains modest for a machine-learning analysis. Selection effects and heterogeneous uncertainties can affect the fitted coefficients. A more complete analysis could include these uncertainties in a hierarchical Bayesian model.}

{An observer-frame analysis that removes the shared luminosity-distance factor directly from the feature space would provide a less model-dependent test of a possible intrinsic prompt--plateau association.}

{The same symbolic-search procedure can also be applied to prompt spectral relations and multimessenger observables.}

\section{Conclusions}

{We applied symbolic regression to identify analytic relations for GRB plateau luminosity. The Dainotti-guided search recovers the classical anticorrelation. In the unrestricted search over 377 feature-subset search spaces, the Dainotti relation remains competitive but ranks within the top $\sim25\%$ rather than first.}

The {highest-ranked screened} symbolic model is
\[
\log L_a \approx 0.855 \log L_{\rm peak} - 0.782 \log T_a^{\rm rest},
\]
with validation RMSE $\approx 0.459$ and $R^2 \approx 0.882$.

An explicit multivariate regression gives a prompt--plateau plane,
\[
\log L_a = 4.718 + 0.776 \log L_{\rm peak} - 0.805 \log T_a^{\rm rest},
\]
with RMSE $\approx 0.535$ and $R^2 \approx 0.824$. {After removing shared luminosity-distance scaling, the relation retains non-negligible fit quality (RMSE $\approx 0.569$, $R^2 \approx 0.62$), which is consistent with a possible intrinsic component.} The corresponding scaling is sublinear,
\[
L_a \propto L_{\rm peak}^{0.78} T_a^{-0.81}.
\]

{An errors-in-variables D'Agostini refit gives consistent slopes, $(0.773\pm0.054)$ for $\log L_{\rm peak}$ and $(-0.808\pm0.076)$ for $\log T_a^{\rm rest}$, with intrinsic scatter $\sigma_{\rm int}=0.512\pm0.040$ dex. The class-dependent fit using the smaller catalog with \texttt{Type} labels gives the same positive prompt-luminosity and negative plateau-time dependence in the larger subclasses. The smaller subclasses are not large enough for a stable class-by-class symbolic search.}

{A post-selection repeated-split check of the selected empirical forms gives the same ordering: the prompt--plateau plane has lower validation MAE than the classical OLS Dainotti relation in all 100 random train-validation splits.}

{The energy diagnostics show that plateau energy is not constant and increases with prompt luminosity and duration. The plateau phase therefore follows a coupled energy--timescale relation rather than a one-dimensional scaling.}
{This energy diagnostic is consistent with earlier Dainotti-relation work in suggesting a prompt/afterglow energy-budget connection, but our distance-controlled analysis shows that part of the raw energy trend is strengthened by shared cosmological scaling.}

{In the restricted Monte Carlo test, the real-data relation is separated from the randomized results ($p_{\rm corr}\approx0.01$). Because the strongest models contain $L_{\rm peak}$, this result represents the recovery of a prompt--plateau association in this sample rather than a completely independent new relation.}

{The two-variable prompt--plateau plane provides a broader description in which the classical Dainotti relation is a one-dimensional projection: plateau duration remains essential, while prompt peak luminosity accounts for much of the additional structure in $\log L_a$. The highest-ranked symbolic expressions repeatedly omit the additional spectral, absorption, and afterglow-shape variables. Within this sample and search setup, the dominant low-dimensional structure is therefore the association among prompt luminosity, plateau duration, and plateau luminosity. Symbolic regression provides an interpretable way to identify, rank, and reconstruct such empirical relations in time-domain astrophysics.}

\section{{Data and Code Availability}}
\label{sec:code_availability}

The catalogs, analysis code, archived stochastic-search outputs, and machine-readable summary tables used in this work are provided in the accompanying Zenodo repository via \dataset[DOI: 10.5281/zenodo.21967404]{https://doi.org/10.5281/zenodo.21967404}. The package includes a Jupyter notebook, deterministic support scripts, and versioned environment specifications. Its default workflow reproduces the deterministic fits, robustness checks, and diagnostic figures from the supplied data and archived search outputs. Documented switches allow the computationally expensive PySR searches and null experiments to be rerun separately, and the README describes the inputs, outputs, software environment, and execution commands.

\appendix

\section{{Exploratory Redshift Prediction}}
\label{sec:redshift}

{As an additional test, we applied symbolic regression to estimate GRB redshifts from the observed quantities. Redshift estimation is more difficult than the plateau-luminosity regression because the luminosity distance, $k$-correction, and rest-frame times all depend on redshift. We therefore divided the 251 GRBs into 188 training objects and 63 test objects before calculating any rest-frame quantity. Using only the training sample, we obtained symbolic relations for the plateau luminosity, peak luminosity, and intrinsic column density. For each test burst, a trial redshift determines the luminosity distance, rest-frame times, and flux correction \((1+z)^{\Gamma-2}\). We varied the trial redshift over \(10^{-4}\le z\le10\) and selected the value that minimizes the weighted squared differences from the three symbolic relations, normalized by their training scatters, together with a weak prior constructed only from observed quantities.}

{For the 63 test GRBs, the symbolic estimate gives a Pearson correlation $R_{\rm P}=0.441$ and RMSE $=1.464$. For comparison, an observed-only random-forest model gives $R_{\rm P}=0.571$ and RMSE $=1.364$ for the same objects. Therefore, the symbolic method does not improve the redshift estimate relative to this baseline. The main difficulty is the degeneracy between redshift and the observed quantities when no direct distance information is available. This degeneracy limits the performance of a low-dimensional symbolic model.}

{For the present data, symbolic regression is more suitable for finding interpretable GRB correlations than for estimating redshift. We include this test to show this limitation of the method. The correlation search in the main analysis is a forward problem, whereas estimating redshift from the observed quantities is a less constrained inverse problem.}

\begin{acknowledgments}
R.M. acknowledges support from the Beijing Natural Science Foundation (IS24021) and the Institute of High Energy Physics, Chinese Academy of Sciences (E32984U810).. {The authors used OpenAI Codex \citep{OpenAICodex2026} for language polishing and as an additional review of all Python code included in the reproducibility package. All AI-assisted changes were reviewed and approved by the authors, who take responsibility for the final content.} We thank the developers of PySR \citep{2023arXiv230501582C} for making symbolic regression accessible to the scientific community.
\end{acknowledgments}

\software{Python, NumPy, pandas, SciPy, scikit-learn, Astropy, Matplotlib, PySR}


\begin{thebibliography}{}
\expandafter\ifx\csname natexlab\endcsname\relax\def\natexlab#1{#1}\fi
\providecommand{\url}[1]{\href{#1}{#1}}

\end{thebibliography}


\begin{thebibliography}{}
\expandafter\ifx\csname natexlab\endcsname\relax\def\natexlab#1{#1}\fi
\providecommand{\url}[1]{\href{#1}{#1}}

\bibitem[{{Andreon} \& {Hurn}(2013)}]{2013SADM....6...15A}
{Andreon}, S., \& {Hurn}, M. 2013, Statistical Analysis and Data Mining: The
  ASA Data Science Journal, 9, 15

\bibitem[{{Berger}(2014)}]{2014ARA&A..52...43B}
{Berger}, E. 2014, \araa, 52, 43

\bibitem[{Bernardini(2015)}]{BERNARDINI201564}
Bernardini, M.~G. 2015, Journal of High Energy Astrophysics, 7, 64, swift 10
  Years of Discovery, a novel approach to Time Domain Astronomy.
\newblock
  \url{https://www.sciencedirect.com/science/article/pii/S221440481500021X}

\bibitem[{{Cao} {et~al.}(2022{\natexlab{a}}){Cao}, {Dainotti}, \&
  {Ratra}}]{2022MNRAS.516.1386C}
{Cao}, S., {Dainotti}, M., \& {Ratra}, B. 2022{\natexlab{a}}, \mnras, 516, 1386

\bibitem[{{Cao} {et~al.}(2022{\natexlab{b}}){Cao}, {Khadka}, \&
  {Ratra}}]{2022MNRAS.510.2928C}
{Cao}, S., {Khadka}, N., \& {Ratra}, B. 2022{\natexlab{b}}, \mnras, 510, 2928

\bibitem[{Cardone {et~al.}(2009)Cardone, Capozziello, \&
  Dainotti}]{10.1111/j.1365-2966.2009.15456.x}
Cardone, V.~F., Capozziello, S., \& Dainotti, M.~G. 2009, Monthly Notices of
  the Royal Astronomical Society, 400, 775.
\newblock \url{https://doi.org/10.1111/j.1365-2966.2009.15456.x}

\bibitem[{Cardone {et~al.}(2010)Cardone, Dainotti, Capozziello, \&
  Willingale}]{10.1111/j.1365-2966.2010.17197.x}
Cardone, V.~F., Dainotti, M.~G., Capozziello, S., \& Willingale, R. 2010,
  Monthly Notices of the Royal Astronomical Society, 408, 1181.
\newblock \url{https://doi.org/10.1111/j.1365-2966.2010.17197.x}

\bibitem[{{Cranmer}(2023)}]{2023arXiv230501582C}
{Cranmer}, M. 2023, arXiv e-prints, arXiv:2305.01582

\bibitem[{{Cranmer} {et~al.}(2020){Cranmer}, {Sanchez-Gonzalez}, {Battaglia},
  {Xu}, {Cranmer}, {Spergel}, \& {Ho}}]{2020arXiv200611287C}
{Cranmer}, M., {Sanchez-Gonzalez}, A., {Battaglia}, P., {et~al.} 2020, arXiv
  e-prints, arXiv:2006.11287

\bibitem[{{D'Agostini}(2005)}]{2005physics..11182D}
{D'Agostini}, G. 2005, arXiv e-prints, physics/0511182

\bibitem[{{Dai} \& {Lu}(1998)}]{1998A&A...333L..87D}
{Dai}, Z.~G., \& {Lu}, T. 1998, \aap, 333, L87

\bibitem[{{Dainotti} {et~al.}(2023){Dainotti}, {Bargiacchi}, {Bogdan},
  {Lenart}, {Iwasaki}, {Capozziello}, {Zhang}, \&
  {Fraija}}]{2023ApJ...951...63D}
{Dainotti}, M.~G., {Bargiacchi}, G., {Bogdan}, M., {et~al.} 2023, \apj, 951, 63

\bibitem[{{Dainotti} {et~al.}(2008){Dainotti}, {Cardone}, \&
  {Capozziello}}]{2008MNRAS.391L..79D}
{Dainotti}, M.~G., {Cardone}, V.~F., \& {Capozziello}, S. 2008, \mnras, 391,
  L79

\bibitem[{{Dainotti} {et~al.}(2011{\natexlab{a}}){Dainotti}, {Fabrizio
  Cardone}, {Capozziello}, {Ostrowski}, \& {Willingale}}]{2011ApJ...730..135D}
{Dainotti}, M.~G., {Fabrizio Cardone}, V., {Capozziello}, S., {Ostrowski}, M.,
  \& {Willingale}, R. 2011{\natexlab{a}}, \apj, 730, 135

\bibitem[{{Dainotti} {et~al.}(2017{\natexlab{a}}){Dainotti}, {Hernandez},
  {Postnikov}, {Nagataki}, {O'brien}, {Willingale}, \&
  {Striegel}}]{2017ApJ...848...88D}
{Dainotti}, M.~G., {Hernandez}, X., {Postnikov}, S., {et~al.}
  2017{\natexlab{a}}, \apj, 848, 88

\bibitem[{Dainotti {et~al.}(2023)Dainotti, Lenart, Chraya, Sarracino, Nagataki,
  Fraija, Capozziello, \& Bogdan}]{Dainotti:2022ked}
Dainotti, M.~G., Lenart, A.~L., Chraya, A., {et~al.} 2023, Mon. Not. Roy.
  Astron. Soc., 518, 2201

\bibitem[{{Dainotti} {et~al.}(2020{\natexlab{a}}){Dainotti}, {Lenart},
  {Sarracino}, {Nagataki}, {Capozziello}, \& {Fraija}}]{2020ApJ...904...97D}
{Dainotti}, M.~G., {Lenart}, A.~{\L}., {Sarracino}, G., {et~al.}
  2020{\natexlab{a}}, \apj, 904, 97

\bibitem[{{Dainotti} {et~al.}(2017{\natexlab{b}}){Dainotti}, {Nagataki},
  {Maeda}, {Postnikov}, \& {Pian}}]{2017A&A...600A..98D}
{Dainotti}, M.~G., {Nagataki}, S., {Maeda}, K., {Postnikov}, S., \& {Pian}, E.
  2017{\natexlab{b}}, \aap, 600, A98

\bibitem[{{Dainotti} {et~al.}(2011{\natexlab{b}}){Dainotti}, {Ostrowski}, \&
  {Willingale}}]{2011MNRAS.418.2202D}
{Dainotti}, M.~G., {Ostrowski}, M., \& {Willingale}, R. 2011{\natexlab{b}},
  \mnras, 418, 2202

\bibitem[{{Dainotti} {et~al.}(2013){Dainotti}, {Petrosian}, {Singal}, \&
  {Ostrowski}}]{2013ApJ...774..157D}
{Dainotti}, M.~G., {Petrosian}, V., {Singal}, J., \& {Ostrowski}, M. 2013,
  \apj, 774, 157

\bibitem[{{Dainotti} {et~al.}(2015){Dainotti}, {Petrosian}, {Willingale},
  {O'Brien}, {Ostrowski}, \& {Nagataki}}]{2015MNRAS.451.3898D}
{Dainotti}, M.~G., {Petrosian}, V., {Willingale}, R., {et~al.} 2015, \mnras,
  451, 3898

\bibitem[{{Dainotti} {et~al.}(2016){Dainotti}, {Postnikov}, {Hernandez}, \&
  {Ostrowski}}]{2016ApJ...825L..20D}
{Dainotti}, M.~G., {Postnikov}, S., {Hernandez}, X., \& {Ostrowski}, M. 2016,
  \apjl, 825, L20

\bibitem[{{Dainotti} {et~al.}(2010){Dainotti}, {Willingale}, {Capozziello},
  {Fabrizio Cardone}, \& {Ostrowski}}]{2010ApJ...722L.215D}
{Dainotti}, M.~G., {Willingale}, R., {Capozziello}, S., {Fabrizio Cardone}, V.,
  \& {Ostrowski}, M. 2010, \apjl, 722, L215

\bibitem[{{Dainotti} {et~al.}(2020{\natexlab{b}}){Dainotti}, {Livermore},
  {Kann}, {Li}, {Oates}, {Yi}, {Zhang}, {Gendre}, {Cenko}, \&
  {Fraija}}]{2020ApJ...905L..26D}
{Dainotti}, M.~G., {Livermore}, S., {Kann}, D.~A., {et~al.} 2020{\natexlab{b}},
  \apjl, 905, L26

\bibitem[{{Dainotti} {et~al.}(2026){Dainotti}, {Lenart}, {De Simone},
  {Giar{\`e}}, {Di Valentino}, {Hartmann}, {Fraija}, {Iwasaki}, \&
  {Lambiase}}]{2026arXiv260318223D}
{Dainotti}, M.~G., {Lenart}, A.~{\L}., {De Simone}, B., {et~al.} 2026, arXiv
  e-prints, arXiv:2603.18223

\bibitem[{{Dall'Osso} {et~al.}(2011){Dall'Osso}, {Stratta}, {Guetta}, {Covino},
  {De Cesare}, \& {Stella}}]{2011A&A...526A.121D}
{Dall'Osso}, S., {Stratta}, G., {Guetta}, D., {et~al.} 2011, \aap, 526, A121

\bibitem[{{D'Avanzo}(2015)}]{2015JHEAp...7...73D}
{D'Avanzo}, P. 2015, Journal of High Energy Astrophysics, 7, 73

\bibitem[{Draper \& Smith(1998)}]{draper1998applied}
Draper, N.~R., \& Smith, H. 1998, Applied regression analysis, Vol. 326 (John
  Wiley \& Sons)

\bibitem[{{Efron} \& {Petrosian}(1992)}]{1992ApJ...399..345E}
{Efron}, B., \& {Petrosian}, V. 1992, \apj, 399, 345

\bibitem[{Efron \& Tibshirani(1994)}]{Efron1994}
Efron, B., \& Tibshirani, R.~J. 1994, An Introduction to the Bootstrap (New
  York: Chapman \& Hall/CRC), doi:10.1201/9780429246593

\bibitem[{{Evans} {et~al.}(2009){Evans}, {Beardmore}, {Page}, {Osborne},
  {O'Brien}, {Willingale}, {Starling}, {Burrows}, {Godet}, {Vetere}, {Racusin},
  {Goad}, {Wiersema}, {Angelini}, {Capalbi}, {Chincarini}, {Gehrels}, {Kennea},
  {Margutti}, {Morris}, {Mountford}, {Pagani}, {Perri}, {Romano}, \&
  {Tanvir}}]{2009MNRAS.397.1177E}
{Evans}, P.~A., {Beardmore}, A.~P., {Page}, K.~L., {et~al.} 2009, \mnras, 397,
  1177

\bibitem[{Favale {et~al.}(2024)Favale, Dainotti, G\'omez-Valent, \&
  Migliaccio}]{Favale:2024lgp}
Favale, A., Dainotti, M.~G., G\'omez-Valent, A., \& Migliaccio, M. 2024, JHEAp,
  44, 323

\bibitem[{{Feigelson} \& {Babu}(2012)}]{2012msma.book.....F}
{Feigelson}, E.~D., \& {Babu}, G.~J. 2012, {Modern Statistical Methods for
  Astronomy}, doi:10.48550/arXiv.1205.2064

\bibitem[{{Gehrels} {et~al.}(2006){Gehrels}, {Norris}, {Barthelmy}, {Granot},
  {Kaneko}, {Kouveliotou}, {Markwardt}, {M{\'e}sz{\'a}ros}, {Nakar}, {Nousek},
  {O'Brien}, {Page}, {Palmer}, {Parsons}, {Roming}, {Sakamoto}, {Sarazin},
  {Schady}, {Stamatikos}, \& {Woosley}}]{2006Natur.444.1044G}
{Gehrels}, N., {Norris}, J.~P., {Barthelmy}, S.~D., {et~al.} 2006, \nat, 444,
  1044

\bibitem[{{Isobe} {et~al.}(1990){Isobe}, {Feigelson}, {Akritas}, \&
  {Babu}}]{1990ApJ...364..104I}
{Isobe}, T., {Feigelson}, E.~D., {Akritas}, M.~G., \& {Babu}, G.~J. 1990, \apj,
  364, 104

\bibitem[{{Jin} {et~al.}(2019){Jin}, {Fu}, {Kang}, {Guo}, \&
  {Guo}}]{2019arXiv191008892J}
{Jin}, Y., {Fu}, W., {Kang}, J., {Guo}, J., \& {Guo}, J. 2019, arXiv e-prints,
  arXiv:1910.08892

\bibitem[{{Kumar} {et~al.}(2008){Kumar}, {Narayan}, \&
  {Johnson}}]{2008MNRAS.388.1729K}
{Kumar}, P., {Narayan}, R., \& {Johnson}, J.~L. 2008, \mnras, 388, 1729

\bibitem[{{Kumar} \& {Zhang}(2015)}]{2015PhR...561....1K}
{Kumar}, P., \& {Zhang}, B. 2015, \physrep, 561, 1

\bibitem[{{Lenart} {et~al.}(2025){Lenart}, {Dainotti}, {Khatiya}, {Bal},
  {Hartmann}, {Fraija}, \& {Zhang}}]{Lenart2025}
{Lenart}, A.~{\L}., {Dainotti}, M.~G., {Khatiya}, N., {et~al.} 2025, Journal of
  High Energy Astrophysics, 47, 100384

\bibitem[{{Liu} {et~al.}(2025){Liu}, {Lu}, {Moradi}, {Yang}, {Zhang}, {Lin}, \&
  {Wang}}]{2025arXiv251204204L}
{Liu}, Y., {Lu}, Y., {Moradi}, R., {et~al.} 2025, arXiv e-prints,
  arXiv:2512.04204

\bibitem[{{L{\"u}} {et~al.}(2015){L{\"u}}, {Zhang}, {Lei}, {Li}, \&
  {Lasky}}]{2015ApJ...805...89L}
{L{\"u}}, H.-J., {Zhang}, B., {Lei}, W.-H., {Li}, Y., \& {Lasky}, P.~D. 2015,
  \apj, 805, 89

\bibitem[{{Makke} \& {Chawla}(2022)}]{2022arXiv221110873M}
{Makke}, N., \& {Chawla}, S. 2022, arXiv e-prints, arXiv:2211.10873

\bibitem[{{M{\'e}sz{\'a}ros}(2006)}]{2006RPPh...69.2259M}
{M{\'e}sz{\'a}ros}, P. 2006, Reports on Progress in Physics, 69, 2259

\bibitem[{{Moradi} {et~al.}(2025){Moradi}, {Wang}, {Rastegarnia},
  {Yorgancioglu}, {Yi}, {Eslam Panah}, \& {Zhang}}]{2025ApJ...993...68M}
{Moradi}, R., {Wang}, Y., {Rastegarnia}, F., {et~al.} 2025, \apj, 993, 68

\bibitem[{{Narendra} {et~al.}(2025){Narendra}, {Dainotti}, {Sarkar}, {Lenart},
  {Bogdan}, {Pollo}, {Zhang}, {Rabeda}, {Petrosian}, \&
  {Iwasaki}}]{2025A&A...698A..92N}
{Narendra}, A., {Dainotti}, M.~G., {Sarkar}, M., {et~al.} 2025, \aap, 698, A92

\bibitem[{{Nousek} {et~al.}(2006){Nousek}, {Kouveliotou}, {Grupe}, {Page},
  {Granot}, {Ramirez-Ruiz}, {Patel}, {Burrows}, {Mangano}, {Barthelmy},
  {Beardmore}, {Campana}, {Capalbi}, {Chincarini}, {Cusumano}, {Falcone},
  {Gehrels}, {Giommi}, {Goad}, {Godet}, {Hurkett}, {Kennea}, {Moretti},
  {O'Brien}, {Osborne}, {Romano}, {Tagliaferri}, \&
  {Wells}}]{2006ApJ...642..389N}
{Nousek}, J.~A., {Kouveliotou}, C., {Grupe}, D., {et~al.} 2006, \apj, 642, 389

\bibitem[{{OpenAI}(2026)}]{OpenAICodex2026}
{OpenAI}. 2026, Codex, \url{https://openai.com/codex/}, , , accessed 2026 July
  3

\bibitem[{{Pavlidou} {et~al.}(2012){Pavlidou}, {Richards}, {Max-Moerbeck},
  {King}, {Pearson}, {Readhead}, {Reeves}, {Stevenson}, {Angelakis},
  {Fuhrmann}, {Zensus}, {Giroletti}, {Reimer}, {Healey}, {Romani}, \&
  {Shaw}}]{2012ApJ...751..149P}
{Pavlidou}, V., {Richards}, J.~L., {Max-Moerbeck}, W., {et~al.} 2012, \apj,
  751, 149

\bibitem[{{Piran}(1999)}]{1999PhR...314..575P}
{Piran}, T. 1999, \physrep, 314, 575

\bibitem[{{Piran}(2004)}]{2004RvMP...76.1143P}
---. 2004, Reviews of Modern Physics, 76, 1143

\bibitem[{{Planck Collaboration} {et~al.}(2020){Planck Collaboration},
  {Aghanim}, {Akrami}, {Ashdown}, {Aumont}, {Baccigalupi}, {Ballardini},
  {Banday}, {Barreiro}, {Bartolo}, {Basak}, {Battye}, {Benabed}, {Bernard},
  {Bersanelli}, {Bielewicz}, {Bock}, {Bond}, {Borrill}, {Bouchet}, {Boulanger},
  {Bucher}, {Burigana}, {Butler}, {Calabrese}, {Cardoso}, {Carron},
  {Challinor}, {Chiang}, {Chluba}, {Colombo}, {Combet}, {Contreras}, {Crill},
  {Cuttaia}, {de Bernardis}, {de Zotti}, {Delabrouille}, {Delouis}, {Di
  Valentino}, {Diego}, {Dor{\'e}}, {Douspis}, {Ducout}, {Dupac}, {Dusini},
  {Efstathiou}, {Elsner}, {En{\ss}lin}, {Eriksen}, {Fantaye}, {Farhang},
  {Fergusson}, {Fernandez-Cobos}, {Finelli}, {Forastieri}, {Frailis},
  {Fraisse}, {Franceschi}, {Frolov}, {Galeotta}, {Galli}, {Ganga},
  {G{\'e}nova-Santos}, {Gerbino}, {Ghosh}, {Gonz{\'a}lez-Nuevo}, {G{\'o}rski},
  {Gratton}, {Gruppuso}, {Gudmundsson}, {Hamann}, {Handley}, {Hansen},
  {Herranz}, {Hildebrandt}, {Hivon}, {Huang}, {Jaffe}, {Jones}, {Karakci},
  {Keih{\"a}nen}, {Keskitalo}, {Kiiveri}, {Kim}, {Kisner}, {Knox},
  {Krachmalnicoff}, {Kunz}, {Kurki-Suonio}, {Lagache}, {Lamarre}, {Lasenby},
  {Lattanzi}, {Lawrence}, {Le Jeune}, {Lemos}, {Lesgourgues}, {Levrier},
  {Lewis}, {Liguori}, {Lilje}, {Lilley}, {Lindholm}, {L{\'o}pez-Caniego},
  {Lubin}, {Ma}, {Mac{\'\i}as-P{\'e}rez}, {Maggio}, {Maino}, {Mandolesi},
  {Mangilli}, {Marcos-Caballero}, {Maris}, {Martin}, {Martinelli},
  {Mart{\'\i}nez-Gonz{\'a}lez}, {Matarrese}, {Mauri}, {McEwen}, {Meinhold},
  {Melchiorri}, {Mennella}, {Migliaccio}, {Millea}, {Mitra},
  {Miville-Desch{\^e}nes}, {Molinari}, {Montier}, {Morgante}, {Moss}, {Natoli},
  {N{\o}rgaard-Nielsen}, {Pagano}, {Paoletti}, {Partridge}, {Patanchon},
  {Peiris}, {Perrotta}, {Pettorino}, {Piacentini}, {Polastri}, {Polenta},
  {Puget}, {Rachen}, {Reinecke}, {Remazeilles}, {Renzi}, {Rocha}, {Rosset},
  {Roudier}, {Rubi{\~n}o-Mart{\'\i}n}, {Ruiz-Granados}, {Salvati}, {Sandri},
  {Savelainen}, {Scott}, {Shellard}, {Sirignano}, {Sirri}, {Spencer},
  {Sunyaev}, {Suur-Uski}, {Tauber}, {Tavagnacco}, {Tenti}, {Toffolatti},
  {Tomasi}, {Trombetti}, {Valenziano}, {Valiviita}, {Van Tent}, {Vibert},
  {Vielva}, {Villa}, {Vittorio}, {Wandelt}, {Wehus}, {White}, {White},
  {Zacchei}, \& {Zonca}}]{2020A&A...641A...6P}
{Planck Collaboration}, {Aghanim}, N., {Akrami}, Y., {et~al.} 2020, \aap, 641,
  A6

\bibitem[{{Postnikov} {et~al.}(2014){Postnikov}, {Dainotti}, {Hernandez}, \&
  {Capozziello}}]{2014ApJ...783..126P}
{Postnikov}, S., {Dainotti}, M.~G., {Hernandez}, X., \& {Capozziello}, S. 2014,
  \apj, 783, 126

\bibitem[{{Rowlinson} {et~al.}(2013){Rowlinson}, {O'Brien}, {Metzger},
  {Tanvir}, \& {Levan}}]{2013MNRAS.430.1061R}
{Rowlinson}, A., {O'Brien}, P.~T., {Metzger}, B.~D., {Tanvir}, N.~R., \&
  {Levan}, A.~J. 2013, \mnras, 430, 1061

\bibitem[{{Sari} {et~al.}(1999){Sari}, {Piran}, \&
  {Halpern}}]{1999ApJ...519L..17S}
{Sari}, R., {Piran}, T., \& {Halpern}, J.~P. 1999, \apjl, 519, L17

\bibitem[{Schmidt \& Lipson(2010)}]{Schmidt2010}
Schmidt, M., \& Lipson, H. 2010, in Genetic Programming Theory and Practice
  VII, ed. R.~Riolo, U.-M. O'Reilly, \& T.~McConaghy (Boston, MA: Springer US),
  73--85.
\newblock \url{https://doi.org/10.1007/978-1-4419-1626-6_5}

\bibitem[{{Srinivasaragavan} {et~al.}(2020){Srinivasaragavan}, {Dainotti},
  {Fraija}, {Hernandez}, {Nagataki}, {Lenart}, {Bowden}, \&
  {Wagner}}]{2020ApJ...903...18S}
{Srinivasaragavan}, G.~P., {Dainotti}, M.~G., {Fraija}, N., {et~al.} 2020,
  \apj, 903, 18

\bibitem[{Taper \& Lele(2004)}]{10.7208/chicago/9780226789583.001.0001}
Taper, M.~L., \& Lele, S.~R. 2004, The Nature of Scientific Evidence:
  Statistical, Philosophical, and Empirical Considerations (University of
  Chicago Press), doi:10.7208/chicago/9780226789583.001.0001.
\newblock \url{https://doi.org/10.7208/chicago/9780226789583.001.0001}

\bibitem[{{Uhm} \& {Zhang}(2014)}]{2014ApJ...789...39U}
{Uhm}, Z.~L., \& {Zhang}, B. 2014, \apj, 789, 39

\bibitem[{{Wadekar} {et~al.}(2023){Wadekar}, {Thiele}, {Hill}, {Pandey},
  {Villaescusa-Navarro}, {Spergel}, {Cranmer}, {Nagai},
  {Angl{\'e}s-Alc{\'a}zar}, {Ho}, \& {Hernquist}}]{2023MNRAS.522.2628W}
{Wadekar}, D., {Thiele}, L., {Hill}, J.~C., {et~al.} 2023, \mnras, 522, 2628

\bibitem[{Wang {et~al.}(2024)Wang, Moradi, \& Li}]{Wang:2024fun}
Wang, Y., Moradi, R., \& Li, L. 2024, Astrophys. J., 974, 89

\bibitem[{{Willmott} \& {Matsuura}(2005)}]{2005ClRes..30...79W}
{Willmott}, C., \& {Matsuura}, K. 2005, Climate Research, 30, 79

\bibitem[{{Yorgancioglu} {et~al.}(2025){Yorgancioglu}, {Saeed}, {Moradi}, \&
  {Wang}}]{2025ApJ...989...60Y}
{Yorgancioglu}, E.~S., {Saeed}, D.~M., {Moradi}, R., \& {Wang}, Y. 2025, \apj,
  989, 60

\bibitem[{{Zhang}(2018)}]{2018pgrb.book.....Z}
{Zhang}, B. 2018, {The Physics of Gamma-Ray Bursts} (Cambridge University
  Press), doi:10.1017/9781139226530

\bibitem[{{Zhang} {et~al.}(2006){Zhang}, {Fan}, {Dyks}, {Kobayashi},
  {M{\'e}sz{\'a}ros}, {Burrows}, {Nousek}, \& {Gehrels}}]{2006ApJ...642..354Z}
{Zhang}, B., {Fan}, Y.~Z., {Dyks}, J., {et~al.} 2006, \apj, 642, 354

\bibitem[{{Zhang} \& {M{\'e}sz{\'a}ros}(2001)}]{2001ApJ...552L..35Z}
{Zhang}, B., \& {M{\'e}sz{\'a}ros}, P. 2001, \apjl, 552, L35

\bibitem[{{Zhang} \& {M{\'e}sz{\'a}ros}(2004)}]{2004IJMPA..19.2385Z}
---. 2004, International Journal of Modern Physics A, 19, 2385

\end{thebibliography}
\bibliographystyle{aasjournal}

\end{document}